\documentclass[twocolumn, deluxetables]{aastex631}
\usepackage{amsmath}
\usepackage[caption=false]{subfig}
\usepackage{multirow}
\usepackage{graphicx}
\usepackage{xcolor}
\usepackage[T1]{fontenc} 
\usepackage{enumitem}
\usepackage{verbatim}

\newcommand{\chandra}{\textit{Chandra}}
\newcommand{\nicer}{\textit{NICER}}

\newcommand{\swift}{\textit{Swift}}
\newcommand{\rosat}{\textit{ROSAT}}

\newcommand{\xmm}{\textit{XMM-Newton}}

\newcommand{\xspec}{\texttt{xspec}}

\accepted{July 3, 2024}

\shorttitle{\xmm\ Spectral Analysis of AT2021ehb}
\shortauthors{Xiang et al.}

\graphicspath{{./}{figures/}}

\begin{document}

\title{Investigating the Mass of the Black Hole and Possible Wind Outflow of the Accretion Disk in the Tidal Disruption Event AT2021ehb}

\correspondingauthor{Xin Xiang}
\email{xinxiang@umich.edu}

\author[0000-0002-7129-4654]{Xin Xiang}
\affiliation{Department of Astronomy, The University of Michigan, 1085 South University Avenue, Ann Arbor, MI 48103, USA}

\author[0000-0003-2869-7682]{Jon M. Miller}
\affiliation{Department of Astronomy, The University of Michigan, 1085 South University Avenue, Ann Arbor, MI 48103, USA}

\author{Abderahmen Zoghbi}
\affiliation{Department of Astronomy, The University of Maryland College Park, MD 20742, USA}
\affiliation{HEASARC, Code 6601, NASA/GSFC, Greenbelt, MD 20771, USA}
\affiliation{CRESST II, NASA Goddard Space Flight Center, Greenbelt, MD 20771, USA}

\author[0000-0003-1621-9392]{Mark T. Reynolds}
\affiliation{Department of Astronomy, The University of Michigan, 1085 South University Avenue, Ann Arbor, MI 48103, USA}
\affiliation{Department of Astronomy, Ohio State University, 140 W. 18th Avenue, Columbus, OH 43210, USA}

\author[0000-0002-5924-4822]{David Bogensberger}
\affiliation{Department of Astronomy, The University of Michigan, 1085 South University Avenue, Ann Arbor, MI 48103, USA}

\author[0000-0002-9589-5235]{Lixin Dai}
\affiliation{Department of Physics, The University of Hong Kong, Pokfulam Road, Hong Kong}

\author[0000-0002-2218-2306]{Paul A. Draghis}
\affiliation{Department of Astronomy, The University of Michigan, 1085 South University Avenue, Ann Arbor, MI 48103, USA}

\author[0000-0002-0210-2276]{Jeremy J. Drake}
\affiliation{Lockheed Martin, 3251 Hanover Street, Palo Alto, CA 94304}

\author{Olivier Godet}
\affiliation{IRAP, Université de Toulouse, CNRS, CNES, 9 Avenue du Colonel Roche, 31028 Toulouse, France}

\author{Jimmy Irwin}
\affiliation{Department of Physics and Astronomy, The University of Alabama, Tuscaloosa, AL 35487, USA}

\author[0000-0002-2666-728X]{M. Coleman Miller}
\affiliation{Department of Astronomy, The University of Maryland College Park, MD 20742, USA}

\author[0000-0001-6350-8168]{Brenna Mockler}
\affiliation{Department of Physics and Astronomy, University of California, Los Angeles, CA 90095, USA}
\affiliation{The Observatories of the Carnegie Institution for Science, Pasadena, CA 91101, USA}

\author[0000-0002-4912-2477]{Richard Saxton}
\affiliation{Telespazio-Vega UK for ESA, Operations Department, European Space Astronomy Centre(ESAC), Villanueva de la Cañada 28692, Madrid, Spain}

\author{Natalie Webb}
\affiliation{IRAP, Université de Toulouse, CNRS, CNES, 9 Avenue du Colonel Roche, 31028 Toulouse, France}

\begin{abstract}
Tidal disruption events (TDEs) can potentially probe low-mass black holes in host galaxies that might not adhere to bulge or stellar-dispersion relationships.  At least initially, TDEs can also reveal super-Eddington accretion.  X-ray spectroscopy can potentially constrain black hole masses, and reveal ionized outflows associated with super-Eddington accretion. Our analysis of \xmm\ X-ray observations of the TDE AT2021ehb, around 300 days post-disruption, reveals a soft spectrum and can be fit with a combination of multi-color disk blackbody and power-law components.  Using two independent disk models with properties suited to TDEs, we estimate a black hole mass at $M \simeq 10^{5.5}~M_{\odot}$, indicating AT2021ehb may expose the elusive low-mass end of the nuclear black hole population.  These models offer simple yet robust characterization; more complicated models are not required, but provide important context and caveats in the limit of moderately sensitive data.  If disk reflection is included, the disk flux is lower and inferred black hole masses are $\sim$ 0.35 dex higher.  Simple wind formulations imply an extremely fast $v_{\mathrm{out}} = -0.2~c$ outflow and obviate a disk continuum component.  Assuming a unity filling factor, such a wind implies an instantaneous mass outflow rate of $\dot{M} \simeq 5~M_{\odot}~{\rm yr}^{-1}$.  Such a high rate suggests that the filling factor for the Ultra Fast Outflow (UFO) must be extremely low, and/or the UFO phase is ephemeral.  We discuss the strengths and limitations of our analysis and avenues for future observations of TDEs.

\end{abstract}

\keywords{Tidal disruption; X-ray transient sources; Supermassive black holes; Time domain astronomy; High energy astrophysics; Accretion}

\section{Introduction} \label{sec:intro}
Stars occasionally orbit massive black holes (MBHs) closely enough to be disrupted by tidal forces in a tidal disruption event (TDE; see  
review by \citealt{Gezari_2021}). TDEs generate multi-wavelength flares with supernova-like luminosities, while the relativistic jets that accompany a subset of them can be observable from cosmological distances \citep{Zauderer_2011,Cenko_2012}. TDEs provide an opportunity to investigate the demography of quiescent low-mass supermassive black holes (SMBHs) and potentially the intermediate-mass black holes (IMBHs) population \citep{Greene_2020}, which could shed light on the formation path of SMBHs through the constraints on the BH occupation fraction into low-mass galaxies \citep{Wang_2004} as well as the BH mass and spin. TDEs can also serve as cosmic laboratories to study the mechanisms of jet formation, strong outflows, and energetics, which impact ejecting gas in low-mass galaxies, potentially quenching the star formation. They also provide a context for studying super-Eddington accretion, which is a possible mechanism to help grow MBH into SMBH \citep{Greene_2020}.

The theoretical concept of TDEs, which originated in the late 1970s \citep{Hills_1975, Rees_1988, Phinney_1989}, was first supported by observational evidence \citep{Bade_1996} captured as soft X-ray flares from the centers of quiescent galaxies in the \rosat\ All-Sky Survey (RASS) in the 1990s. TDEs are primarily characterized by a flux that diminishes over time loosely following a $t^{-5/3}$ decay pattern \citep{Phinney_1989, Evans_1989}, and by thermal emission, exhibiting a characteristic temperature of $T_{\rm eff} \sim 10^6 K$ for X-ray bright TDEs \citep{Sazonov_2021}. The majority of the X-ray spectra of TDEs are soft and can be modeled with a modified blackbody approximating the disk emission \citep{Mummery_2021}, which can in principle be used to infer the temperature and the size of the X-ray emitting region, hence the estimation of the black hole mass \citep{Mummery_Wevers_Saxton_Pasham_2023}. \textbf{}

The capabilities of observatories such as \chandra, \xmm , \swift, as well as wide-field surveys across different wavelengths including UV and optical surveys (e.g., PTF, iPTF, ZTF, ASAS-SN, Pan-STARRS, SDSS, and ATLAS), has led to an increasing number of TDE discoveries \citep{Gezari_2021}. However, there are still ongoing debates regarding the nature and evolution of the accretion flow that occurs after a star is disrupted. The larger size of the inferred blackbody radius of the UV/optical TDEs thermal component and the lower effective temperature $T_{\rm eff} \sim 10^4 K$ compared with the X-ray TDE candidates implies a larger structure that can potentially be produced by winds or an outflow \citep{Miller_2015, Kara_2018} and their characteristics dependence on reprocessing of the emission \citep{Dai_2018, Thomsen_2022}, or self-collision of the stream debris around the black hole \citep{Bonnerot_2017}.

AT2021ehb was initially detected by the Zwicky Transient Facility (ZTF) on 2021 March 1 (MJD 59274) with a g-band magnitude of $19.2$ \citep{Munoz-Arancibia_2021}. Subsequently, \swift\ conducted high-cadence monitoring during the rising phase of AT2021ehb, leading to its classification as a tidal disruption event (TDE) on 2021 March 26 based on characteristics such as its extended rise time, broad spectral features, bright UV color \citep{Gezari_2021}, and X-rays detection \citep{Yao_2021}. AT2021ehb has a redshift of $z = 0.018$ \citep{Yao_2022b}, corresponding to a luminosity distance of 78 Mpc assuming a standard $\Lambda$CDM cosmology with default values of the parameters in \xspec\ \citep{Arnaud_1996} ($\Omega_M = 0.27$, $\Omega_\Lambda = 0.73$, and $H_0 = 70$ km s$^{-1}$ Mpc$^{-1}$).

The host galaxy of AT2021ehb has a total stellar mass of $M_* \sim 10^{10} M_\odot$ and a BH mass of $M_{BH} \sim 10^{7} M_\odot$ based on galactic scaling relationships \citep{Yao_2022a}.  The multi-wavelength emission of AT2021ehb over its first 430 days, spanning X-ray, UV, optical, and radio wavelengths, was studied in detail by \citet{Yao_2022a}. Their X-ray spectra show potential evidence of relativistic disk reflection with a broad iron K line and tentative suggestions of an aspherical accretion flow geometry. The bolometric luminosity reached up to 6\% of the Eddington luminosity. Notably, during these first 430 days, the X-ray spectrum exhibits a transition from a soft to a hard state, followed by a rapid X-ray flux drop to an intermediate state within 3 days. These transitions of the state are possibly due to the formation of a magnetically-dominated corona and the impact of thermal-viscous instability in the inner accretion flow. The potential outflow originating from the shock provides valuable insights into the formation of accretion disks and coronae. While \citet{Yao_2022a} focused on the broad multi-wavelength characteristics and potential outflow mechanisms, our study specifically aims to analyze the X-ray spectral properties and explore the presence of Ultra-fast Outflows (UFOs) in AT2021ehb during the intermediate state (phase E1 in \citealt{Yao_2022a} where the first drops in X-ray luminosity occur). This provides a more detailed investigation into the accretion dynamics and the properties of the outflows.

In this paper, we present an X-ray spectral analysis of AT2021ehb using the spectra obtained with \xmm\ in early 2022. We also present the optical light curves based on monitoring observations made with the \swift\ \citep{Gehrels_2004}. The observations and data reduction are outlined in section \ref{sec:data}.  We present the results of X-ray spectral analysis in Section \ref{sec:results}, where we find a black hole mass, approximately $M \sim 10^{5.5} M_\odot$, which is smaller than the prediction based on galaxy scaling relations. An alternative indication from the data is the presence of Ultra-fast Outflows (UFOs). The implications, discussions, and summary of our key findings are presented in Section \ref{sec:discussion}.

\section{Observation and Data Reduction} \label{sec:data}

\subsection{The Swift/UVOT light curve}

\begin{figure}
\includegraphics[width=0.47\textwidth]{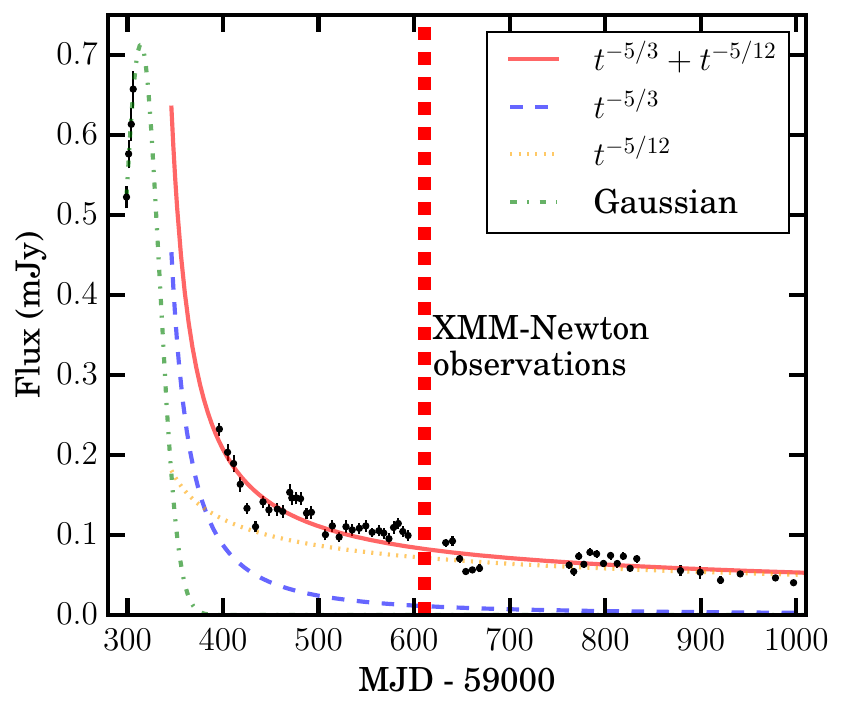}
\caption{The \swift/UVOT light curve of AT2021ehb using M2 filter. The host contribution has been removed using the host galaxy SED models in \citet{Yao_2022a}}. The red curve is the fitting result of a typical TDE power-law decay that combines a $t^{-5/3}$ (blue dashed curve) and a $t^{-5/12}$ (yellow dotted line). The green dash-dotted line is the fitting result of a Gaussian function for the data before the peak.
\label{fig: LC}
\end{figure}

The \swift\ X-ray Telescope (XRT, \citealt{Roming_2005}) and Ultra-violet Optical Telescope (UVOT, \citealt{Burrows_2005}) data were reduced using the tools in HEASARC version 6.31.1, and the latest calibration files available through the standard public release.   The full set of UV filters (UVW1, UVM2, UVW2, with central wavelengths of 260~nm, 220~nm, and 190~nm, respectively) appear to be available in each individual exposure.  However, the M2 filter has the smallest transmission leak at its long wavelength edge, making it the most robust trace of the UV emission.  For this reason, we have elected to restrict our characterization of the UV light curve to the M2 filter.  

All individual M2 exposures were given astrometric corrections, and subdivided exposures (in a single filter) were summed.  Source fluxes were extracted in circular regions with a radius of 3'' using \texttt{uvotmaghist}, while backgrounds were extracted from nearby source-free regions of equal size.  Flux levels were then corrected for coincidence losses, using the coincidence loss figures calculated by \texttt{uvotmaghist}.

\citet{Yao_2022a} constructed the pre-TDE host galaxy SED using the photometric data from SDSS, the Two Micron All-Sky Survey, and the AllWISE catalog. The best-fit SED model, shown in Figure 5 of \citet{Yao_2022a}, predicts an AB magnitude of approximately 22.5 in the M2 band. We converted this value into flux density and subtracted it from each data point of the M2 light curve. Figure \ref{fig: LC} shows the host subtracted M2 light curve of AT2021ehb. The peak of the M2 light curve occurred during Sun occultation and cannot be robustly determined due to the lack of data between MJD 59306 -- 59396. A simple Gaussian fit for the early rising data indicates the peak occurs around MJD $59314^{+36}_{-8}$. The lower bound is a hard limit determined by the last data point before the peak. The upper error is determined by the best fit of the power-law decay after the peak. The early decay is rapid, broadly consistent with the $F \propto t^{-5/3}$ profile expected if the mass accretion rate directly follows from the fallback rate.  However, the late-time decay is clearly flatter, consistent with the $F \propto t^{-5/12}$ profile expected if the evolution is driven by a standard geometrically thin, and optically thick, accretion disk \citep{Lodato_2011}. 

\subsection{\xmm}
We obtained six epochs of observations with \xmm\ in late January and early February of 2022 with observation IDs of 084014201 (Obs 1), 084014301 (Obs 2), 084014401 (Obs 3), 084014901 (Obs 4), 084014601 (Obs 5), 0840141001 (Obs 6). The observations were taken with the European Photon Imaging Camera (EPIC) and the Reflection Grating Spectrometer (RGS). Since EPIC is optimized for a broader energy range in X-rays than RGS, and the pn instrument generally has better sensitivity than MOS1 and MOS2, we only analyze the pn data. A log of the \xmm\ EPIC-pn observations is given in Table \ref{table:observationlog}.

\begin{deluxetable}{ccccc}
\tablecaption{Log of six epochs of observations of AT2021ehb with \xmm\ EPIC-PN with Obs IDs of 084014201 (Obs 1), 084014301 (Obs 2), 084014401 (Obs 3), 084014901 (Obs 4), 084014601 (Obs 5), 0840141001 (Obs 6).}
\label{table:observationlog}
\tablewidth{0pt}
\tablehead{
\colhead{obsID} & \colhead{MJD} & \colhead{Exposure} & \colhead{$\delta t$} & \colhead{0.5-10 keV flux} \\
& & (ks) & (days) & ($10^{-12}~\mathrm{erg~s^{-1}~cm^{-2}}$)
}

\startdata
    084014201 & 59606 & 15.3 & +292 & $4.00^{+0.04}_{-0.04}$\\
    084014301 & 59608 & 15.3 & +294 & $4.17^{+0.03}_{-0.03}$ \\
    084014401 & 59610 & 15.3 & +296 & $3.98^{+0.04}_{-0.04}$ \\
    084014901 & 59612 & 15.3 & +298 & $2.95^{+0.03}_{-0.03}$ \\
    084014601 & 59614 & 15.3 & +300 & $3.04^{+0.04}_{-0.04}$ \\
    0840141001 & 59616 & 15.3 & +302 & $2.77^{+0.04}_{-0.04}$ \\ \hline
\enddata
\tablecomments{$\delta t$ is the days that have passed after the peak. The 0.5-10 keV fluxes are computed using the \texttt{cflux} function in \texttt{xspec} for the best-fit model (a1). See details of model (a1) in Figure \ref{fig: fitall} and Table \ref{table:continum}.}
\end{deluxetable}

The data were reduced using the \xmm\ Standard Analysis Software (SAS version 20.0.0). Calibrated event lists for the EPIC-pn were produced using the \texttt{epproc} task. The source and background spectra were generated using \texttt{evselect} task with patterns that correspond to single and double events (PATTERN <= 4). RMFs and ARFs were generated using \texttt{rmfgen} and \texttt{arfgen} tools. The \texttt{ftgrouppha} task in HEASOFT version 6.30.1 was used to bin the spectrum using the optimal binning scheme \cite{Kaastra_Bleeker_2016}.

\section{Results and Analysis} \label{sec:results}
\subsection{Modeling the \xmm \ X-ray spectra}
In this section, we present a spectral analysis of the six epochs of observation data in the 0.3--8.0 keV range separately fitting with multiple models. The fits presented below are performed with \xspec\ version 12.12.1 \citep{Arnaud_1996}.  We used the \texttt{vern} cross sections \citep{Verner_1996} and the \texttt{wilm} abundances \citep{Wilms_2000}. Cash statistics \citep{Cash_1979} were used and uncertainties are $1\sigma$ ranges for parameters. 

For all the models described below, we included galactic absorption using the \texttt{tbabs} model \citep{Wilms_2000}, with the hydrogen-equivalent column density of $N_H = 9.95 \times 10^{20}~\mathrm{cm}^{-2}$ \citep{Collaboration_2016}.  The host galactic absorption is not included in the model because its column density is negligibly small both in our data and prior treatments of CCD spectra of AT2021ehb  \citep{Yao_2022a}. The x-axes for all the spectral plots are shifted by a factor of $1.018$ to allow for viewing in the source frame. For illustrative purposes, all plotted data bins are grouped into a maximum of 10 bins reaching the 5 sigma significance criterion (\texttt{"setplot rebin 5, 10"} in \texttt{xspec}). 

\begin{figure}
\includegraphics[width=0.47\textwidth]{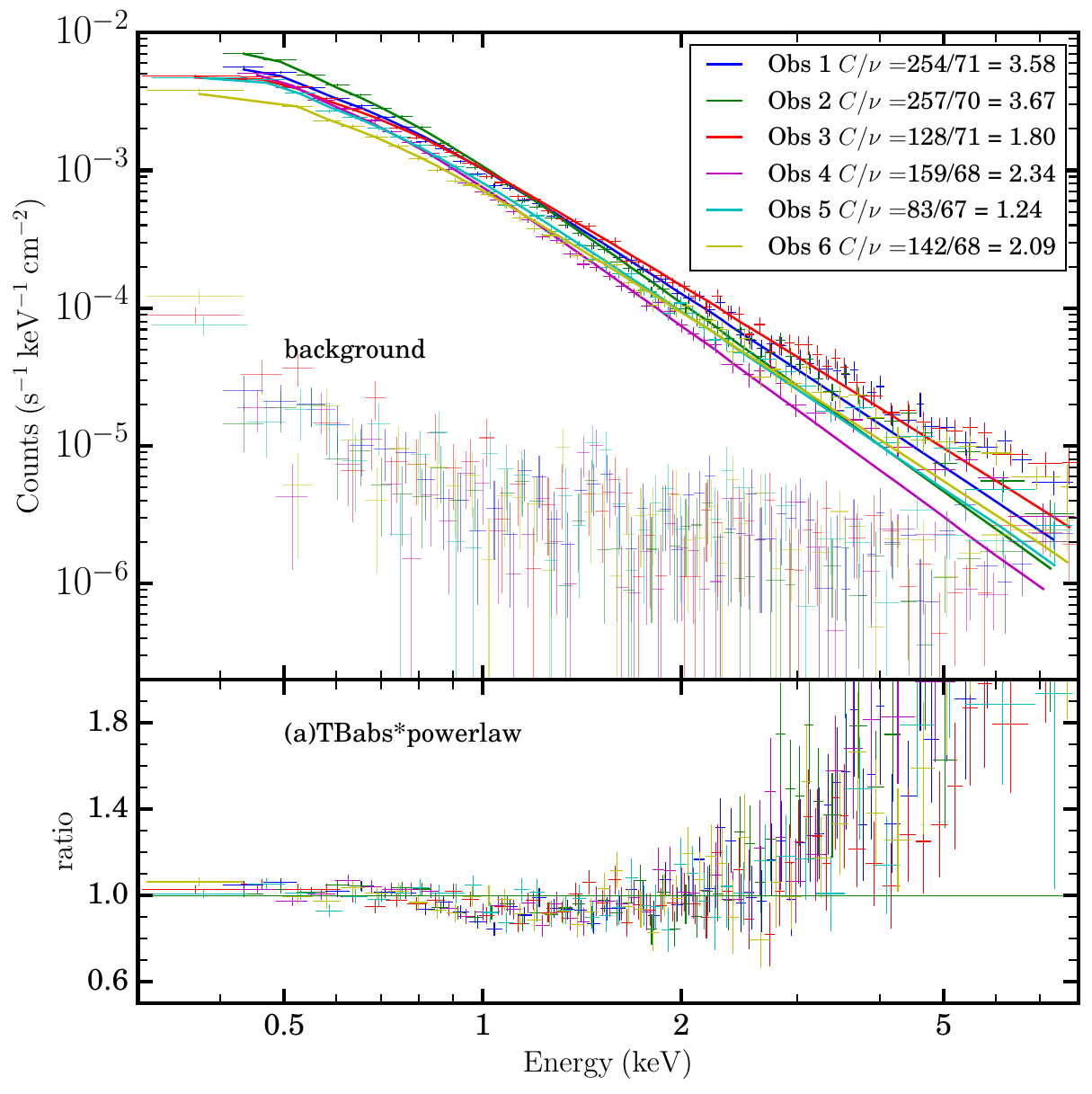}
\caption{
The \xmm\ spectra of AT2021ehb, and their corresponding backgrounds. The fit with a single power-law is shown as curves with the corresponding color.  The models are forced to very steep, likely unphysical spectral indices (see table \ref{table:continum} for fit results), and fail to fit the data above 3~keV.  The spectra clearly require a thermal component at low energy.}
\label{fig: fita}
\end{figure}

\begin{figure*}
\includegraphics[width=\textwidth]{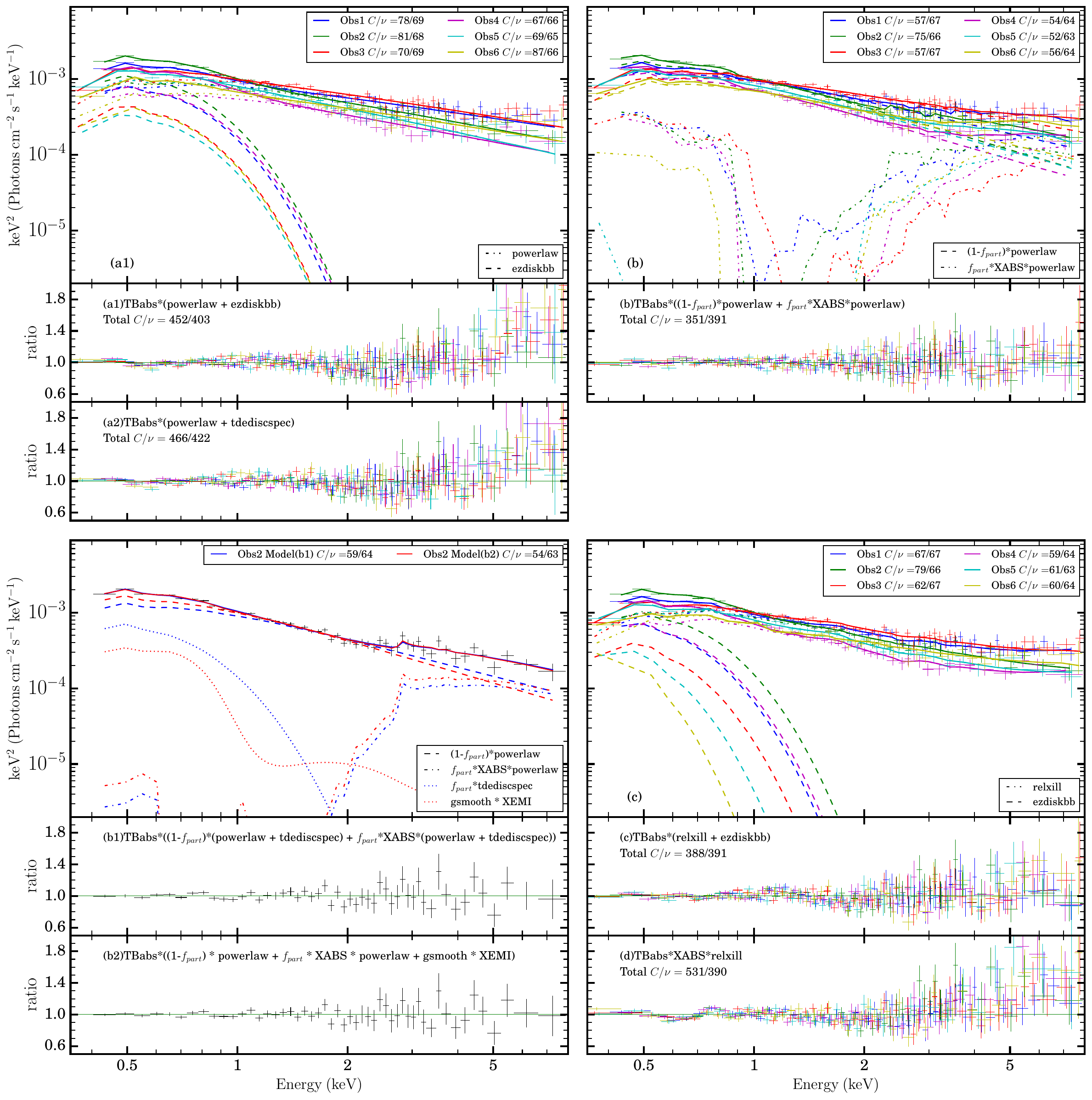}
\caption{
The \xmm\ spectra of AT2021ehb, fit with several plausible models detailed in Tables \ref{table:continum}, \ref{table:bcefall}, and \ref{table:d1d2}. The solid lines show the best-fit curves while the dotted, dashed, and dot-dashed lines are the breakdown of each component of the best-fit models. Adjacent panels show the data/model ratio from each model.   Upper left: phenomenological continuum models consisting of \texttt{powerlaw + ezdiskbb} (model a1) and \texttt{powerlaw + tdediscspec} (model a2). Panel (a1) shows the model (a1) fit with six epochs of the observations.  Upper right: a model including partially covering ionized absorption via a dedicated XSTAR-derived table model \texttt{XABS} (model b).  Lower left: fits for only observation 2 data with dedicated XSTAR-derived table models including absorption \texttt{XABS} and \texttt{tdediscspec} (model b1), and both absorption \texttt{XABS} and smoothed re-emission \texttt{XEMI} (model b2).  Lower right: fits made including relativistic disk reflection via \texttt{relxill} (model c) and with reflection modified by absorption (model d). $C/\nu$ is the ratio of Cash-statistics and the degree of freedom.}
\label{fig: fitall}
\end{figure*}

\subsubsection{Continuum Modeling}
We started modeling the spectrum with a single power-law and obtained an averaged photon index of $\Gamma \sim 3.30$ (see table \ref{table:continum} for details of best-fit values), which is softer than most AGNs ($\Gamma \sim 1.8$ is typical for Seyferts \citealt{Nandra_2007}), in agreement with the claim that TDEs are intrinsically softer than AGNs \citep{Auchettl_2018}. In Figure \ref{fig: fita}, we show the six epochs of \xmm\ spectra fit with the model \texttt{Tbabs*powerlaw}, hereafter model (a). The residuals are most significant beyond 2 keV as a hard excess, potentially indicating the existence of a distinct soft X-ray component such as an additional disk blackbody.

We consider two different disk blackbody models, \texttt{ezdiskbb} \citep{Zimmerman_2005} and \texttt{tdediscspec} \citep{Mummery_2021}, as the additional components.

The \texttt{ezdiskbb} model incorporates a multiple blackbody disk with a zero-torque inner boundary condition. The model assumes that the disk is optically thick and it radiates as a modified blackbody, with a radial temperature profile that has a peak temperature $T_{\mathrm{max}}$ close to the inner radius. The best-fit result with \texttt{TBabs*(powerlaw + ezdiskbb)}, hereafter model (a1), is depicted in the upper left panel of Figure \ref{fig: fitall}, with detailed best-fit parameters presented in Table \ref{table:continum}. The fit for Obs 3 gives the optimal performance with Cash statistics ($C$) of 70 for 69 degrees of freedom ($\nu$), whereas Obs 6 achieves a less satisfactory fit with $C/\nu = 87/66 = 1.32$. Aggregating data from all observations results in a good fit with $C/\nu = 452/403 = 1.12$. For all observations averaged, the peak disk temperature is approximately $0.12~\mathrm{keV}$, aligning with the values of the blackbody temperatures for TDEs with detected X-ray components (ranging from 0.02-0.13 keV \citealt{Gezari_2021}). The disk's inner radius can be deduced from the normalization factor, which is defined by the equation
\begin{equation} \label{eq: Norm}
  \mathrm{Norm} = f^{-4} \left(\frac{R_{\mathrm{in}}}{1\times10^3~\mathrm{m}}\right)^2 \left(\frac{D}{1\times10^4~\mathrm{pc}}\right)^{-2} \cos i 
\end{equation}
where $f$ is the temperature-dependent color correction factor due to the opacity effect \citep{Done_2012}, $R_{\mathrm{in}}$ is the inner radius of the disk, $D$ is the distance to the source, and $i$ is the system inclination. 

For a disk temperature of $T = 0.12~\mathrm{keV}$, the color corrected factor is 2.1, as derived from the relation $f \sim (72/T_{\mathrm{keV}})^{1/9}$ \citep{Done_2012}. This factor may increase due to the temperature decline in the outer disk annulus. We adopt a color correction factor of $f = 2.3$, which is typical for TDE disk temperatures \citep{Mummery_2021}. An inclination angle of $60^\circ$, the average in three dimensions, is assumed. We then calculate the black hole mass estimates, designated as $M^{a=0}_{\mathrm{BH}}$ for a non-spinning black hole (spin $a = 0$) and $M^{a=1}_{\mathrm{BH}}$ for a maximally spinning black hole (spin $a = 1$). These calculations are based on setting the derived inner disk radius ($R_{\mathrm{in}}$) equal to the radius of the Innermost Stable Circular Orbit ($R_\mathrm{ISCO}$) appropriate for a prograde accretion disk at the respective spin values. The value of black hole mass estimation for each observation is presented in Table \ref{table:continum}.  For all the observations averaged, the mean black hole masses are $\log \Bar{M}^{a=0}_{\mathrm{BH}} = 4.99^{+0.04}_{-0.04}~M_\odot$ and $\log \Bar{M}^{a=1}_{\mathrm{BH}} = 5.77^{+0.04}_{-0.04}~M_\odot$, positioning it at the low mass end of the galaxy/supermassive black hole (SMBH) relationship. Assuming an inclination angle of 0 degrees would set the lower mass limit for the black hole at $\log \Bar{M}^{a=0}_{\mathrm{BH, min}} = 4.84^{+0.04}_{-0.04}~M_\odot$ and $\log \Bar{M}^{a=1}_{\mathrm{BH, min}} = 5.62^{+0.04}_{-0.04}~M_\odot$, respectively.

An alternative model we consider is \texttt{tdediscspec}, a recent TDE disk model created by \cite{Mummery_2021}. Unlike traditional disk models, \texttt{tdediscspec} does not presuppose a steady-state configuration for the radial profile, making it more representative of the dynamic nature of TDEs. 

The only assumption of the model is that each disk radius emits like a color-corrected blackbody and there exists a maximum temperature $T_\mathrm{p}$ within the disk at radius $R_\mathrm{p}$. The flux calculated by this model depends only on the hottest temperature in the disk, and not on contributions from other disk regions. This is important as TDE disks represent a class of accretion disks that are evolving. This model represents a more agnostic approach to modeling the TDE spectrum. 

The residuals from the best-fit using the model \texttt{TBabs*(powerlaw + tdediscspec)}, subsequently referred to as (a2), is depicted in the lower upper left section of Figure \ref{fig: fitall}. The complete set of parameters for this fit is detailed in Table \ref{table:continum}. The fit statistics are very close to that of model (a1), with Obs 3 giving the most favorable fit at $C/\nu = 70/68 = 1.03$, and Obs 6 showing the least favorable fit at $C/\nu = 85/65 = 1.31$. When considering all observations collectively, the overall fit statistic stands at $C/\nu = 466/397 = 1.17$. 

Within this model, the constant $\gamma$ depends on the presumed inclination angle of the disk and the disk inner boundary conditions. The value of $\gamma$ spans from $1/2$ to $3/2$, where $\gamma = 1/2$ corresponds to a disk with vanishing ISCO stress observed precisely face-on. The best-fit values of $\gamma$ for all six observations prefer the highest value of $3/2$, indicating a finite ISCO stress disk observed edge-on, in agreement with late-time X-ray TDEs \citep{Mummery_2020}. With the finite ISCO stress, the temperature generally peaks at the inner boundary of the disk \citep{Balbus_2014}. Hence, we can employ the same methodology to estimate the mass of the black hole by equating $R_\mathrm{p} \cos{i}$ with $R_\mathrm{ISCO}$, while assuming specific values for black hole spin ($a = 0$ and $1$) and inclination ($i = 60^o$). The values of the black hole mass estimation for all the observations are detailed in Table \ref{table:continum}. The mean black hole masses are $\log \Bar{M}^{a=1}_{\mathrm{BH}} = 5.82^{+0.005}_{-0.005}~M_\odot$ and $\log \Bar{M}^{a=0}_{\mathrm{BH}} = 5.04^{+0.005}_{-0.005}~M_\odot$, which largely align with model (a1). Furthermore, even without specifying the black hole spin and inclination values, it is possible to estimate the black hole mass. Assuming a uniform spin and inclination distribution, \cite{Mummery_Wevers_Saxton_Pasham_2023} calibrated a radius-to-mass conversion factor $X$ with an average value of $\Bar{X} \sim 4.9^{+7.1}_{-3.0}$, defined by:
\begin{equation}
    \left(\frac{M_{BH, R_\mathrm{p}}}{10^6~M_\odot}\right) = X \left(\frac{R_\mathrm{p}}{10^{12}~\mathrm{cm}}\right),
\end{equation}
According to this method, the black hole mass, for all the observations averaged, is $\log \Bar{M}_{\mathrm{BH, R_p}} = 5.53^{+0.39}_{-0.42}~M_\odot$, which widely agrees with our previous estimation.

The residual plots for both models (a1) and (a2) reveal persistent unmodeled positive residuals at $5-7~\mathrm{keV}$, as well as potential emission and absorption lines around $3~\mathrm{keV}$ and above $7~\mathrm{keV}$. These might be explained by disk reflection features, such as relativistic iron lines or blueshifted photoionized absorption lines. Additionally, while models (a1) and (a2) provide a simple yet robust estimation of the black hole mass if the peak radius is an accurate prediction of the $R_{\mathrm{ISCO}}$, we still need to consider the potential caveats associated with these estimations. One major concern is the suppression of X-ray emission due to reprocessing by an ionized medium, which can cause the radius inferred from thermal soft X-ray emission in TDEs to be much smaller than expected and change over time \citep{Guolo_2023}. To address these issues, we proceeded to explore the photoionization absorption models to describe the data without requiring a disk that might result in an unphysical radius, as detailed below.

\begin{deluxetable*}{cclllllll} 
\label{table:continum}
\tablecaption{Best-fit parameters for continuum models of six \xmm \ EPIC/pn observations of AT2021ehb. For consistency, a galactic absorption component is included via \texttt{TBabs}, with a fixed hydrogen-equivalent column density of $N_H = 9.95 \times 10^{20}~\mathrm{cm}^{-2}$.  Both model (a1) and model (a2) infer a low mass black hole of $M_\mathrm{BH}\sim10^{5-6} M_\odot$ if we assume the inclination of the disk being $i = 60^o$, the average angle in 3D, and the color correction factor of 2.3. The mean black hole masses (for all observation averaged) estimated by both model (a1) and (a2), for scenarios of a non-spinning ($a=0$) and a maximally spinning ($a=1$) black hole, are $\Bar{M}^{a=0}_{\mathrm{BH}} = 10^{5.0}~M_\odot$ and $\Bar{M}^{a=1}_{\mathrm{BH}} = 10^{5.8}~M_\odot$ respectively. If we assume a uniform spin and inclination distribution, the black hole mass computed using the radius-to-mass conversion \citep{Mummery_Wevers_Saxton_Pasham_2023} is $\log \Bar{M}_{\mathrm{BH}, R_\mathrm{p}} \sim 5.5^{+0.4}_{-0.4}~M_\odot$. In model \texttt{tdediscspec}, $R_\mathrm{p}$ refers to the disk radius where the maximum temperature, $T_\mathrm{p}$, is observed.}
\tablewidth{0pt}
\tablehead{
\colhead{Models} & \colhead{Components} & \colhead{Parameters} & \colhead{Obs 1} & \colhead{Obs 2} & \colhead{Obs 3} & \colhead{Obs 4} & \colhead{Obs 5} & \colhead{Obs 6}
}
\startdata
    (a) & \texttt{powerlaw} & $\Gamma$ & $3.22^{+0.02}_{-0.01}$ & $3.51^{+0.02}_{-0.01}$ & $3.03^{+0.01}_{-0.01}$ &  $3.56^{+0.02}_{-0.02}$ & $3.31^{+0.02}_{-0.02}$ & $3.14^{+0.02}_{-0.02}$\\
    & & Norm ($\times 10^{-3}$) & $1.26^{+0.01}_{-0.01}$ & $1.32^{+0.01}_{-0.01}$ & $1.26^{+0.01}_{-0.01}$ & $0.93^{+0.01}_{-0.01}$ & $1.00^{+0.01}_{-0.01}$ & $0.87^{+0.01}_{-0.01}$ \\
    & -- & $C/\nu$ & $254/71$ & $257/70$ & $128/71$  & $160/68$ & $83/67$ & $143/68$\\
    & & & $= 3.58$&$= 3.67$&$= 1.80$&$= 2.34$&$= 1.24$&$= 2.09$ \\ \hline
    (a1) & \texttt{powerlaw} & $\Gamma$ & $2.70^{+0.04}_{-0.04}$ & $2.90^{+0.05}_{-0.05}$ & $2.76^{+0.04}_{-0.04}$ & $2.92^{+0.07}_{-0.07}$ & $3.08^{+0.07}_{-0.07}$ & $2.77^{+0.06}_{-0.06}$\\
    & & Norm ($\times 10^{-3}$) & $0.95^{+0.03}_{-0.03}$ & $0.93^{+0.03}_{-0.03}$ & $1.09^{+0.03}_{-0.03}$ & $0.65^{+0.03}_{-0.03}$ & $0.89^{+0.04}_{-0.04}$ & $0.72^{+0.03}_{-0.03}$\\
    & \texttt{ezdiskbb} & $T_{\mathrm{max}}$ (keV) & $0.120^{+0.004}_{-0.004}$ & $0.121^{+0.003}_{-0.003}$ & $0.114^{+0.006}_{-0.007}$ & $0.123^{+0.003}_{-0.003}$ & $0.114^{+0.007}_{-0.009}$ & $0.113^{+0.006}_{-0.006}$\\
    & & Norm & $272^{+44}_{-36}$  & $348^{+40}_{-35}$ & $196^{+61}_{-42}$ & $224^{+32}_{-27}$ & $146^{+65}_{-46}$ & $194^{+57}_{-41}$ \\ 
    & -- & $\log M^{a=0}_{\mathrm{BH}}~(M_\odot)$ & $5.04^{+0.03}_{-0.03}$ & $5.09^{+0.02}_{-0.02}$ & $4.97^{+0.06}_{-0.05}$ & $4.99^{+0.03}_{-0.03}$ & $4.90^{+0.08}_{-0.08}$ & $4.96^{+0.06}_{-0.05}$ \\
    & -- & $\log M^{a=1}_{\mathrm{BH}}~(M_\odot)$ & $5.81^{+0.03}_{-0.03}$ & $5.87^{+0.02}_{-0.02}$  & $5.74^{+0.06}_{-0.05}$ & $5.77^{+0.03}_{-0.03}$ & $5.68^{+0.08}_{-0.08}$ & $5.74^{+0.06}_{-0.05}$\\ 
    & -- & $C/\nu$ & $78/69$ & $81/68$ & $70/69$ & $67/66$ & $69/65$ & $87/66$\\
    & & & $= 1.13$&$= 1.19$&$= 1.01$&$= 1.02$&$= 1.06$&$= 1.32$ \\ \hline
    (a2) & \texttt{powerlaw} & $\Gamma$ & $2.78^{+0.01}_{-0.02}$ & $2.95^{+0.01}_{-0.03}$ & $2.68^{+0.01}_{-0.02}$ & $3.10^{+0.02}_{-0.02}$ & $2.90^{+0.02}_{-0.02}$ & $2.65^{+0.02}_{-0.02}$\\
    & & Norm ($\times 10^{-3}$) & $0.99^{+0.01}_{-0.01}$ & $0.96^{+0.01}_{-0.01}$ & $1.03^{+0.01}_{-0.01}$ & $0.73^{+0.004}_{-0.01}$ & $0.79^{+0.01}_{-0.01}$ & $0.66^{+0.01}_{-0.01}$ \\
    & \texttt{tdediscspec} & $R_p$ ($\times 10^{10}$ cm)& $7.19^{+0.10}_{-0.06}$ & $8.50^{+0.07}_{-0.06}$ & $6.33^{+0.06}_{-0.01}$ & $6.75^{+0.11}_{-0.06}$ & $6.53^{+0.06}_{-0.10}$ & $6.30^{+0.06}_{-0.11}$\\
    & & $T_p$ ($\times 10^{5}$ K) & $11.39^{+0.04}_{-0.04}$ & $11.53^{+0.03}_{-0.03}$ & $11.44^{+0.04}_{-0.06}$ & $11.42^{+0.05}_{-0.04}$ & $11.38^{+0.04}_{-0.06}$ & $11.40^{+0.04}_{-0.06}$\\
    & & $\gamma$ & $1.49^{+0.01}_{-0.02}$ & $1.50^{+0\dagger}_{-0.01}$ & $1.50^{+0\dagger}_{-0.02}$ & $1.50^{+0\dagger}_{-0.02}$ & $1.50^{+0\dagger}_{-0.02}$ & $1.50^{+0\dagger}_{-0.01}$\\
    & -- & $\log M^{a=0}_{\mathrm{BH}}~(M_\odot)$ & $5.061^{+0.006}_{-0.004}$ & $5.133^{+0.004}_{-0.003}$ & $5.005^{+0.004}_{-0.007}$ & $5.033^{+0.007}_{-0.004}$ & $5.019^{+0.004}_{-0.008}$ & $5.003^{+0.004}_{-0.007}$ \\
    & -- & $\log M^{a=1}_{\mathrm{BH}}~(M_\odot)$ & $5.839^{+0.006}_{-0.004}$ & $5.911^{+0.004}_{-0.003}$  & $5.783^{+0.004}_{-0.007}$ & $5.811^{+0.007}_{-0.004}$ & $5.797^{+0.004}_{-0.008}$ & $5.781^{+0.004}_{-0.007}$\\ 
    & -- & $\log M_{\mathrm{BH, R_p}}~(M_\odot)$ & $5.55^{+0.39}_{-0.41}$ & $5.62^{+0.39}_{-0.41}$  & $5.49^{+0.39}_{-0.42}$ & $5.52^{+0.40}_{-0.42}$ & $5.51^{+0.39}_{-0.42}$ & $5.49^{+0.39}_{-0.42}$\\ 
    & -- & $C/\nu$ & $82/68$ & $81/67$ & $70/68$ & $76/65$ & $72/64$ & $85/65$\\
    & & & $= 1.21$&$= 1.21$&$= 1.03$&$= 1.17$&$= 1.12$&$= 1.31$ \\
\enddata
\tablecomments{ (a)\texttt{TBabs*powerlaw}; (a1)\texttt{TBabs*(powerlaw + ezdiskbb)}; \\(a2)\texttt{TBabs*(powerlaw + tdediscspec)}\\ $^*$ The parameters cannot be constrained, or they are insensitive to the fit statistics. \\$^\dagger$ The parameters pegged at hard limit
}
\end{deluxetable*}
\subsubsection{Photoionization Modeling} \label{sec:photmodel}
We explore alternative modifications to the single powerlaw model (a) by considering the addition of absorption features. The hard excess we encountered in model (a) can also indicate a strong absorption component around 1 keV with a harder photon index of the powerlaw component. First, similar to \cite{Miller_2022a}, instead of using \texttt{zxipcf}, which has a very coarse sampling of ionization space, we add a partial absorber to the continuum using a high-resolution XSTAR \citep{Kallman_2001} table model. The table models were generated using the \texttt{xstar2xspec} function of XSTAR, which is a wrapper that generates a table of individual XSTAR runs with specific combinations of parameters. 

\begin{figure}
\includegraphics[width=0.47\textwidth]{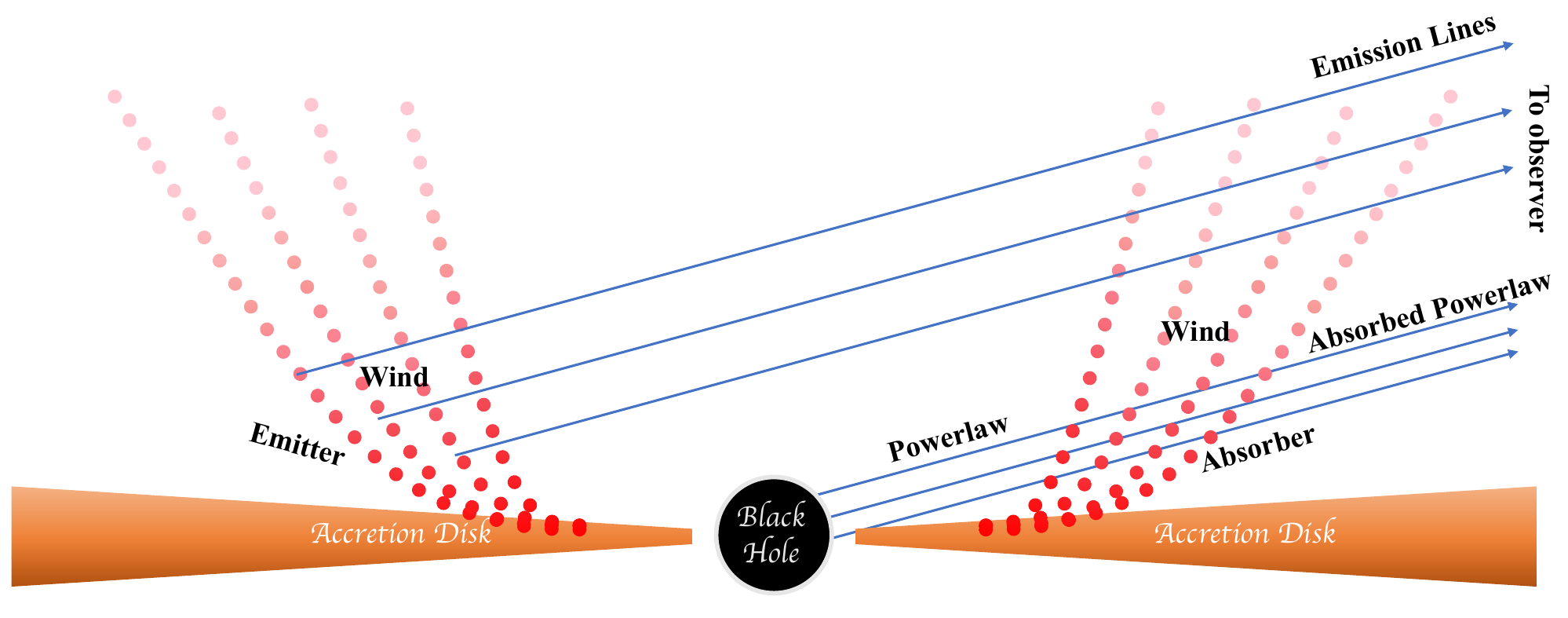}
\caption{Simple schematic diagram of a partial covering absorber and emitter of clumping wind. The absorption of the emitter from the higher region of the wind is neglected.}
\label{fig: geo}
\end{figure}

The tables that we generated have 10,000 spectral bins from 0.1--20 keV. We assumed parameters that are appropriate for Seyfert owing to the similarity of the wind outflow observed in TDEs \citep{Miller_2015, Kara_2018} and Seyfert warm absorbers \citep{Blustin_2005, Laha_2016}, and growing evidence of Seyfert-like broad line regions in TDEs. These included a gas number density of $n = 10^8~\mathrm{cm}^{-3}$, a turbulent velocity of $v_{\mathrm{turb}} = 300~\mathrm{km}~\mathrm{s}^{-1}$, and a luminosity of $L = 6\times10^{44}~\mathrm{erg~s^{-1}}$ between 13.6 eV to 13.6 keV (1--1000 Ry). The input spectral form consists of a T = 25,000 K blackbody and $\Gamma = 1.7$ powerlaw, with the powerlaw artificially absorbed to 0 below 0.3 keV. The resultant tables sample the ionization parameter $1 \leq \log{\xi} \leq 6$ with 100 grid points and the equivalent hydrogen column density $1.0\times 10^{20}~\mathrm{cm}^{-2} \leq {N_H}\leq 6.0\times10^{23}~\mathrm{cm}^{-2}$ with 64 grid points. We assume that the absorber (denoted as \texttt{XABS}) partially covers a fraction ($f_{\mathrm{part}}$) of the source, while the remaining ($1 - f_{\mathrm{part}}$) of the spectrum is seen directly. Figure \ref{fig: geo} depicts the schematic geometry of the absorber and the emitter (the emitter is not included in the current model, see the later discussion on model (b1)). The best-fit model of \texttt{TBabs*((1-$f_{\mathrm{part}}$)*powerlaw+$f_{\mathrm{part}}$*XABS*powerlaw)}, hereafter (b), improves the overall fit statistics to $C/\nu = 351/391 = 0.90$ and is shown in the upper right panel of Figure \ref{fig: fitall}. The best-fit parameters are detailed in Table \ref{table:bcefall}.  Similar to \citep{Danehkar_2018}, we will use the following conventions for velocity and redshift:
\begin{itemize}[]
  \item $z = 0.018$ defines the rest frame of AT2021ehb.
  \item $z_{\mathrm{obs}}$ is the observed redshift of a spectral feature in our reference frame.
  \item $z_{\mathrm{out}} = (1+z_{\mathrm{obs}}) / (1+z) - 1$ gives the redshift of an outflow in the rest frame of AT2021ehb.
  \item $v_{\mathrm{out}} = c((1+z_{\mathrm{out}})^2-1) / ((1+z_{\mathrm{out}})^2+1)$, where $c$ is the speed of light, gives the velocity of an outflow in the rest frame of AT2021ehb.
\end{itemize}
Among the observations, Obs 2 presents the least optimal fit with $C/\nu = 75/66 = 1.14$, which will be discussed in greater detail below, whereas the fit statistics for the remaining observations are around $C/\nu = 0.85$.
We note that all the observations have best-fit outflow velocities higher than $0.03 c$, suggesting an ultra-fast outflow (UFO; a highly ionized absorber with outflow velocities in the range of $0.03\sim 0.3 c$ \citealt{Tombesi_2010}), with high confidence of $P_F > 99.99\%$ for Obs 1, Obs 2, Obs3, Obs 4, and Obs 6, and $P_F > 94.75$ for Obs 5 to exclude fits with zero outflow velocity. All observations also prefer including a non-negligible column density (models with a fixed minimum column density of $N_H = 1.0\times 10^{20}~\mathrm{cm}^{-2}$ are excluded with confidence exceeding 99.99\%). The highly blueshifted Ne VIII, Ne IX, and Fe L lines are the main contributors to the absorption features around 1 keV, which are significant due to their high column density around  $N_H = 5.0\times 10^{23}~\mathrm{cm}^{-2}$. This dense cluster of absorption lines creates a broad dip around 1 keV.

While the partial covering wind absorption model provides an adequate statistical fit overall, Observation 2 stands out due to the largest unmodeled emission features around 3 keV (this can be best seen in the lower left panel of Figure \ref{fig: fitall} without overplotting with other observation), making it the least optimal among the observed datasets, as previously indicated. To address this, we explored two additional models for Obs 2. The first model, hereafter (b1), incorporates an extra \texttt{tdediscspec} blackbody component for an improved continuum representation. The second model, hereafter (b2), introduces Gaussian-smoothed emission features by adding the \texttt{gsmooth*XEMI} with the column of the emitter and the absorber tied together, where \texttt{XEMI} is the XSTAR table model for emission features utilizing the same inputs as \texttt{XABS}. The redshift of the emitter is set to be the same as the host. The best-fit results for (b1) and (b2) are shown in Table \ref{table:d1d2} and the lower left panel of Figure \ref{fig: fitall}.

Model (b1) improves the cash statistic to 59 for 64 degrees of freedom. Upon performing the fit, the constant $\gamma$ in \texttt{tdediscspec} can not be well constrained and is statistically insensitive, suggesting that with the presence of a wind absorption component, the data alone cannot definitively determine whether the disk has finite ISCO stress and whether we are observing edge on or not. Hence, we freeze this parameter to $\gamma = 1.5$. The best-fit value of the peak radius is higher compared with the model (a2), resulting in a slightly higher BH mass estimation. The wind absorber is outflowing faster, is less ionized, and has a lower column density, which is expected because the unabsorbed disk component enhances the soft-X-ray band, making the wind absorb more low-energy photons from the powerlaw. The redshift of the wind outflow absorber $z_{\mathrm{out}} = -0.35^{+0.02}_{-0.02}$ indicates an outflow velocity of $v_{\mathrm{out}} = -0.42^{+0.03}_{-0.03}~c$, which is outside the range of even the most extreme UFOs \citep{Tombesi_2010} decreasing the physical plausibility of this model. 

Model (b2) improves the Cash statistic to 54 for 63 degrees of freedom. The best-fit results are shown in Table \ref{table:d1d2}. The best-fit parameters of the absorber are very close to those of model (b1) but have a larger partial covering factor. The parameter $\sigma$ of \texttt{gsmooth} is the line broadening at $\epsilon = 6$ keV. Its best-fit result of $\sigma = 1.68$ indicates a Keplerian velocity for the emitter $v_{\mathrm{kepl}} = \sigma / (\sigma + \epsilon) = 0.22 c$. This velocity is a lower bound, as it could be higher when considering the disk inclination. Assuming the presence of a disk wind, a standard disk is likely to have formed, extending out to at least $R_{\mathrm{in}} = GM / v_{\mathrm{kepl}}^2 \sim 21~r_g$, where $r_g = GM/c^2$ is the gravitational radius. The escape velocity at $R_{\mathrm{in}}$ can be calculated by $v_{\mathrm{esc}} = \sqrt{2} v_{\mathrm{kepl}} = 0.3~c < v_{\mathrm{out}}$. The outflow velocity is larger than the escape velocity, making it possible to escape and form a wind.

Notably, the ionization parameter of the absorber is higher than that of the emitter. This difference in ionization parameters may be attributed to their varying radial positions relative to the central engine. The gas in the absorber is seen directly from where it is launched, potentially closer to the central engine, resulting in a higher ionization parameter. On the other hand, the emitter represents gas located at various distances around the black hole, which averages out over a larger radius and leads to a lower ionization parameter. This distinction offers insight into the complex dynamics and structure of the gas flows in the vicinity of the black hole.
 
From the lower left panel of Figure \ref{fig: fitall}, we see that the smoothed emitter component in model (b2) closely resembles the disk component seen in model (b1). The emission features in these models are characterized by a broad emission peak at around 0.7 keV, which is notably wide and mimics a disk blackbody emission in model (b1). Also, it is worth noting that the outflow velocity suggested by the absorber in these models is excessively high for the wind, casting doubts on the physical plausibility of this scenario. Furthermore, we experimented with fitting model (b1), as well as \texttt{ezdiskbb} in replacement of \texttt{tdediscspec}, across all the other 5 observations. However, the main parameters of the blackbody model could not be statistically constrained. Thus, given the current data quality, it appears the sensitivity is insufficient to simultaneously accommodate both disk blackbody and outflow scenarios. Moving forward, we will treat the two scenarios as competing models and their statistical preference and physical interpretation will be discussed in session \ref{sec: AIC} and \ref{sec:discussion}.

\begin{deluxetable*}{cclllllll}
\tablecaption{Best-fit parameters for photoionization and reflection models of six \xmm\ observations of AT2021ehb. We include the galactic absorption \texttt{TBabs} with the hydrogen-equivalent column density of $N_H = 9.95 \times 10^{20}~\mathrm{cm}^{-2}$.}
\label{table:bcefall}
\tablewidth{0pt}
\tablehead{
\colhead{Models} & \colhead{Components} & \colhead{Parameters} & \colhead{Obs 1} & \colhead{Obs 2} & \colhead{Obs 3} & \colhead{Obs 4} & \colhead{Obs 5} & \colhead{Obs 6}
}
\startdata
    (b) & \texttt{powerlaw} & $\Gamma$ & $3.10^{+0.03}_{-0.03}$ & $3.45^{+0.04}_{-0.03}$ & $2.84^{+0.03}_{-0.04}$ & $3.42^{+0.05}_{-0.04}$ & $3.35^{+0.02}_{-0.03}$ & $3.12^{+0.04}_{-0.03}$\\
    & & Norm ($\times 10^{-3}$) & $3.44^{+0.38}_{-0.52}$ & $3.47^{+0.28}_{-0.38}$ & $2.15^{+0.21}_{-0.16}$ & $4.89^{+1.00}_{-0.66}$ & $3.07^{+0.85}_{-0.46}$ & $3.47^{+0.49}_{-0.54}$\\
    & \texttt{XABS} & $N_H$ ($\times 10^{22}~\rm cm^{-2}$) & $45.1^{+1.6}_{-2.2}$ & $17.5^{+7.1}_{-1.2}$ & $59.3^{+0.1\dagger}_{-1.6}$ & $54.5^{+3.0}_{-4.7}$ & $41.5^{+13.4}_{-11.0}$ & $57.6^{+2.4 \dagger}_{-6.3}$\\
    & & $\rm \log{\xi} \ (\rm erg\ cm\ s^{-1})$& $3.49^{+0.02}_{-0.04}$ & $3.11^{+0.04}_{-0.04}$ & $3.66^{+0.02}_{-0.05}$ & $3.50^{+0.04}_{-0.02}$ & $3.36^{+0.05}_{-0.18}$ & $3.54^{+0.05}_{-0.14}$\\
    & & $f_{\mathrm{part}}$ & $0.67^{+0.03}_{-0.04}$ & $0.65^{+0.02}_{-0.05}$ & $0.48^{+0.08}_{-0.04}$ & $0.83^{+0.03}_{-0.02}$ & $0.68^{+0.06}_{-0.07}$ & $0.76^{+0.03}_{-0.04}$\\ 
    & & $z_{\mathrm{obs}}$ & $-0.102^{+0.016}_{-0.012}$ & $-0.193^{+0.019}_{-0.020}$ & $-0.270^{+0.021}_{-0.024}$ & $-0.193^{+0.024}_{-0.019}$ & $-0.064^{+0.039}_{-0.038}$ & $-0.087^{+0.014}_{-0.016}$ \\
    & -- & $v_{\mathrm{out}}$ ($c$) & $-0.125^{0.017}_{-0.013}$ & $-0.228^{+0.022}_{-0.024}$ &$-0.321^{+0.026}_{-0.030}$ & $-0.228^{+0.028}_{-0.022}$ & $-0.084^{+0.040}_{-0.041}$ & $-0.108^{+0.015}_{-0.017}$\\ 
    & -- & $C/\nu$ & $57/67$ & $75/66$ & $57/67$ & $54/64$ & $52/63$ & $56/64$\\
    & & & $= 0.85$&$= 1.14$&$= 0.85$&$= 0.84$&$= 0.83$&$= 0.88$ \\ \hline
    (c) & \texttt{relxill} & $\Gamma$ & $2.52^{+0.03}_{-0.03}$ & $2.80^{+0.06}_{-0.04}$ & $2.59^{+0.04}_{-0.02}$ & $2.62^{+0.04}_{-0.04}$ & $2.79^{+0.05}_{-0.06}$ & $2.78^{+0.03}_{-0.03}$ \\
    & & $\log{\xi} \ (\rm erg\ cm\ s^{-1})$ & $3.33^{+0.08}_{-0.06}$ & $3.30^{+0.22}_{-0.24}$ & $3.32^{+0.12}_{-0.08}$ & $3.30^{+0.09}_{-0.04}$ & $3.33^{+0.13}_{-0.08}$ & $1.70^{+0.08}_{-0.11}$  \\
    & & Norm  ($\times 10^{-5}$) & $0.45^{+0.28}_{-0.12}$ & $1.14^{+0.26}_{-0.20}$ & $0.84^{+0.15}_{-0.13}$ & $0.22^{+0.18}_{-0.09}$ & $0.66^{+0.23}_{-0.22}$ & $1.07^{+0.04}_{-0.06}$\\ 
    & & $R_f$ & $3.03^{+4.19}_{-1.43}$ & $0.81^{+0.57}_{-0.39}$ & $1.34^{+0.87}_{-0.50}$ & $5.75^{+4.20}_{-2.36}$ & $2.28^{+3.16}_{-0.73}$ & $2.66^{+0.39}_{-0.37}$ \\
    & \texttt{ezdiskbb} & $T_{\mathrm{max}}$ (keV) & $0.100^{+0.003}_{-0.003}$ & $0.112^{+0.005}_{-0.004}$ & $0.089^{+0.004}_{-0.004}$ & $0.104^{+0.004}_{-0.003}$ & $0.073^{+0.005}_{-0.005}$ & $0.056^{+0.004}_{-0.004}$\\
    & & Norm & $650^{+130}_{-99}$  & $462^{+35}_{-46}$ & $721^{+374}_{-155}$ & $529^{+89}_{-110}$ & $2191^{+1242}_{-797}$ & $14677^{+12306}_{-4614}$ \\ 
    & -- & $\log M^{a=0}_{\mathrm{BH}}~(M_\odot)$ & $5.23^{+0.04}_{-0.04}$ & $5.15^{+0.02}_{-0.03}$ & $5.25^{+0.09}_{-0.05}$ & $5.18^{+0.03}_{-0.05}$ & $5.49^{+0.10}_{-0.10}$ & $5.90^{+0.13}_{-0.08}$ \\
    & -- & $\log M^{a=1}_{\mathrm{BH}}~(M_\odot)$ & $6.00^{+0.04}_{-0.04}$ & $5.93^{+0.02}_{-0.03}$  & $6.03^{+0.09}_{-0.05}$ & $5.96^{+0.03}_{-0.05}$ & $6.27^{+0.10}_{-0.10}$ & $6.68^{+0.13}_{-0.08}$\\
    & -- & $C/\nu$ & $67/67$ & $79/66$ & $62/67$ & $61/64$ & $59/63$ & $60/64$\\
    & & & $= 1.00$&$= 1.20$&$= 0.93$&$= 0.95$&$= 0.94$&$= 0.94$ \\ \hline
    (d) & \texttt{relxill} & $\Gamma$ & $3.01^{+0.02}_{-0.02}$ & $3.27^{+0.02}_{-0.02}$ & $2.96^{+0.01}_{-0.02}$ & $3.31^{+0.02}_{-0.01}$ & $3.25^{+0.03}_{-0.02}$ & $2.98^{+0.01}_{-0.02}$ \\
    & & Norm  ($\times 10^{-5}$) & $2.90^{+0.17}_{-0.31}$ & $3.79^{+0.26}_{-0.18}$ & $2.38^{+0.10}_{-0.09}$ & $3.31^{+0.20}_{0.53}$ & $2.95^{+0.17}_{-0.12}$ & $1.74^{+0.06}_{-0.06}$\\ 
    & & $\rm \log{\xi} \ (\rm erg\ cm\ s^{-1})$ & $0^-$ & $0.12^{+0.10}_{-0.11}$ & $0^-$ & $0^-$ & $0^-$ & $0^-$ \\
    & & $R_f$ & $4.91^{+0.34}_{-0.42}$ & $4.09^{+0.68}_{-1.23}$ & $3.23^{+0.41}_{-0.31}$ & $4.97^{+0.35}_{-0.47}$ & $1.01^{+0.30}_{-0.51}$ & $2.77^{+0.32}_{-0.56}$\\
    & \texttt{XABS} & $N_H$ ($\times 10^{22}~\rm cm^{-2}$) & $48.1^{+0.7}_{-1.2}$ & $6.7^{+14.4}_{-2.9}$ & $0.02^{+0.01}_{-0.01}$ & $48.0^{+12.0\dagger}_{-17.3}$ & $0.98^{+1.43}_{-0.63}$ & $1.3^{+0.8}_{-0.6}$\\
    & & $\rm \log{\xi} \ (\rm erg\ cm\ s^{-1})$ & $4.79^{+0.34}_{-0.32}$ & $4.18^{+0.22}_{-0.11}$ & $0.33^{+0.31}_{-0.27}$ & $5.32^{+0.68\dagger}_{-0.48}$ & $3.84^{+0.20}_{-0.22}$  &$3.55^{+0.14}_{-0.13}$\\
    & & $z_{\mathrm{obs}}$ & $0.0^{+0\dagger}_{-0.02}$ & $-0.008^{+0.008\dagger}_{-0.015}$ & $-0.15^{+0.02}_{-0.01}$ & $-0.002^{+0.002\dagger}_{-0.018}$ & $-0.00^{+0\dagger}_{-0.04}$ & $-0.24^{+0.02}_{-0.02}$ \\
    & -- & $v_{\mathrm{out}}$ ($c$) & $-0.018^{+0}_{-0.020}$ & $-0.026^{+0.008}_{-0.015}$ &$-0.178^{+0.023}_{-0.011}$ & $-0.020^{+0.002}_{-0.018}$ & $-0.018^{+0}_{-0.041}$ & $-0.284^{+0.024}_{-0.024}$\\ 
    & -- & $C/\nu$ & $94/67$ & $121/65$ & $72/67$ & $80/64$ & $74/63$ & $90/64$\\
    & & & $= 1.40$&$= 1.86$&$= 1.07$&$= 1.25$&$= 1.17$&$= 1.41$ \\
\enddata
\tablecomments{(b)\texttt{TBabs*((1-$f_{\mathrm{part}}$)*powerlaw+$f_{\mathrm{part}}$*XABS*powerlaw)}; \\ 
    (c)\texttt{TBabs*(relxill+ezdiskbb)}; (d)\texttt{TBabs*XABS*relxill}.\\ $^*$ The parameter cannot be constrained, or they are insensitive to the fit statistics. \\ $^-$ The parameter is frozen
}
\end{deluxetable*}

\begin{deluxetable}{lcc}
\tablecaption{Best-fit parameters for only observation 2 data with a dedicated XSTAR-derived table model including absorption and \texttt{tdediscspec} (b1), and both absorption and smoothed re-emission (b2). We include the galactic absorption \texttt{TBabs} with the hydrogen-equivalent column density of $N_H = 9.95 \times 10^{20}~\mathrm{cm}^{-2}$.}
\label{table:d1d2}
\tablewidth{0pt}
\tablehead{
\colhead{Parameters} & \colhead{(b1)} & \colhead{(b2)}
}
\startdata
&  \texttt{tdediscspec}  &\\
$T_{\mathrm{p}}$ ($\times 10^5 K$) & $11.33^{+0.44}_{-0.48}$ & ...\\
$R_{\mathrm{p}}$ ($\times 10^{10}~\mathrm{cm}$) & $10.48^{+1.53}_{-1.23}$ & ...\\ 
$\gamma$ & $1.5^*$ & ... \\ 
$\log M^{a=0}_{\mathrm{BH}}~(M_\odot)$ & $5.22^{+0.04}_{-0.04}$ & ... \\
$\log M^{a=1}_{\mathrm{BH}}~(M_\odot)$ & $6.00^{+0.04}_{-0.04}$ & ... \\ 
$\log M_{\mathrm{BH, R_p}}~(M_\odot)$ & $5.71^{+0.45}_{-0.47}$ & ... \\ \hline
& \texttt{powerlaw} & \texttt{powerlaw} \\
$\Gamma$ & $3.24^{+0.10}_{-0.10}$ & $3.43^{+0.06}_{-0.06}$ \\
Norm  ($\times 10^{-3}$)& $2.23^{+0.56}_{-0.42}$ & $3.01^{+0.41}_{-0.37}$ \\\hline
& \texttt{XABS} & \texttt{XABS} \\
$N_H$ ($\times 10^{22}~\rm cm^{-2}$)& $6.50^{+2.39}_{-1.02}$ & $6.52^{+1.88}_{-1.19}$ \\
$\rm \log{\xi} \ (\rm erg\ cm\ s^{-1})$ & $1.94^{+0.36}_{-0.28}$ & $1.95^{+0.51}_{-0.27}$ \\
$z_{\mathrm{obs}}$& $-0.35^{+0.02}_{-0.02}$ & $-0.35^{+0.02}_{-0.01}$\\ \hline
$v_{\mathrm{out}}$ ($c$)& $-0.42^{+0.03}_{-0.03}$ & $-0.42^{+0.03}_{-0.03}$ \\ \hline
$f_{\mathrm{part}}$ & $0.52^{+0.07}_{-0.06}$ &  $0.61^{+0.05}_{-0.05}$\\ \hline
&  & \texttt{XEMI} \\
$\rm \log{\xi} \ (\rm erg\ cm\ s^{-1})$ & ... & $-2.43^{+0.53}_{-0.16}$ \\
Norm  ($\times 10^{-3}$) & ... & $6.73^{+5.64}_{-5.21}$ \\ \hline
& & \texttt{gsmooth} \\
$\sigma \rm \ (keV)$ & ... & $1.68^{+0.21}_{-0.18}$ \\ \hline
$C/\nu$ & $59/64$ & $54/63$\\
&$= 0.92$&$= 0.86$\\\hline
\enddata
\tablecomments{\\(b1)\texttt{TBabs*((1-$f_{\mathrm{part}}$)*(powerlaw + tdediscspec) +$f_{\mathrm{part}}$*XABS*(powerlaw+tdediscspec)}; (b2)\texttt{TBabs*((1-$f_{\mathrm{part}}$)*powerlaw\\+$f_{\mathrm{part}}$*XABS*powerlaw+gsmooth*XEMI)};\\
    $^*$The parameters cannot be constrained, or they are insensitive to the fit statistics. \\
    $^\dagger$The parameters pegged at hard limit}
\end{deluxetable}

\subsubsection{Reflection Modeling}
To explore the possibility that a fraction of the non-power-law emission features are part of a disk reflection spectrum, we replaced the power-law component in the model (a1) with the self-consistent \texttt{relxill} model v.2.2 \citep{Dauser_2014, García_2014}. The outer disk radius ($R_{\mathrm{out}}$) is fixed at a fiducial value of $999~r_g$. The inner disk radius ($R_{\mathrm{in}}$) is fixed at the ${\mathrm{ISCO}}$. The emissivity in \texttt{relxill} is modeled by an empirical broken power law, which changes at the break radius $R_{br}$ from $r^{q_1}$ to $r^{q_2}$, where $q_1$, and $q_2$ are the emissivity indexes. We set both $q_1$ and $q_2$ to 3, making the parameter $R_{br}$ obsolete, to align with the expectations for a standard flat and thin disk \citep{Shakura_1973} with an isotropic source of emission, where the energy dissipation of the disk decreases with radius as $r^{-3}$ due to the effect of flux dilution and limb darkening. The redshift parameter is set to be the host redshift. The spin parameter $a$ and the disk inclination are fixed at 0.5 and an average value of $60^o$. The power-law energy cutoff $E_{\mathrm{cut}}$ and the iron abundance of the accretion disk $A_{\mathrm{Fe}}$ are fixed at $300$ keV and solar abundance. The other parameters in \texttt{relxill} are free, including the power-law photon index of the incident spectrum $\Gamma$, the ionization of the accretion disk $\xi$, the reflection fraction $R_f$ defined as the ratio of the intensity of the primary source irradiating the disk and the intensity directly going to infinity (More details are in \citealt{Dauser_2016}), and the normalization parameter $Norm$. 

The best-fit result for the model (c), \texttt{TBabs*(relxill+ezdiskbb)} is detailed in Table \ref{table:bcefall} and illustrated in the lower right panel of Figure \ref{fig: fitall}. The fit statistics for individual observations vary, with the best performance observed for Obs 3 at $C/\nu = 62/67 = 0.93$ and the least optimal for Obs 2 at $C/\nu = 79/66 = 1.2$. Overall, this model yields a total Cash statistic of $C/\nu = 389/391 = 0.99$, offering a noticeable improvement over model (a1). From this model, the black hole mass computed from $Norm$ across all observations, when averaged, amounts to $\log \Bar{M}^{a=0}_{\mathrm{BH}} = 5.37^{+0.07}_{-0.06}~M_\odot$ and $\log \Bar{M}^{a=1}_{\mathrm{BH}} = 6.14^{+0.07}_{-0.06}~M_\odot$. These values are approximately 0.35 dex higher than those derived from model (a1). The reason for this increase in estimated masses is attributed to the lower $T_{\mathrm{max}}$ in the model (c), which results in a higher $Norm$, particularly notable in Obs 6, where $T_{\mathrm{max}} \sim 0.06$, about half the value of the best-fit values from other observations.

Next, we attempted to model the absorbed reflection by combining the \texttt{relxill} model with \texttt{XABS}. The best-fit model \texttt{TBabs*XABS*relxill}, hereafter (d), resulted in a total Cash statistics of $C/\nu = 531/390 = 1.36$, with the best performance for Obs 3 at $C/\nu = 72/67 = 1.07$ and the least optimal for Obs 2 at $C/\nu  = 121/65 = 1.86$. The data fails to effectively constrain the ionization parameter $\xi$ of the reflection model due to the excessive complexity of the model except for Obs 2. The parameters of the absorber are less consistent throughout different observations than in model (b). 

\begin{deluxetable*}{ccccccccc}
    \tablecaption{The values of Akaike information criterion (AIC) and Bayesian information criterion (BIC) for each model fit. The top section presents the summation of the values across all observations, while the bottom section is specific to observation 2 (Obs 2). $\sigma_C$ is the sigma level of the C-statistics compared to degrees of freedom. Ideally, $|\sigma_C|$ should fall within the unity range, making model (c) a viable candidate when considering all observations collectively. A positive $\sigma_C$ value might suggest overfitting in the case of model (b).
    $\Delta[AIC]_p$ and $\Delta[BIC]_p$ are the differences between AIC and BIC of the current model compared to the previous simpler model, from which we can clearly see that the simplest model (a) have no support against models (a1) and (a2), whereas models (b1) and (b2) show considerable more support over model (b) for Obs 2.
    $\Delta[AIC]_{\mathrm{min}}$ and $\Delta[BIC]_{\mathrm{min}}$ -- the difference in AIC and BIC compared to the model with the minimum AIC and BIC -- identify model (b) as having the lowest values of the criterion, with model (c) or (a1) as the next best alternatives.
    Models with $\Delta [AIC]_{\min}$ and $\Delta [BIC]_{\min}$ values greater than 10 for a single observation are less favored, narrowing our focus to a series of photon ionization models (b, b1, b2), and disk emission and reflection models (c, a1, a2) for further consideration.}
    \label{tab:AIC}
    \tablewidth{0pt}
    \tablehead{Model&AIC   &BIC & $\sigma_C$ &$\Delta[AIC]_p$&$\Delta[BIC]_p$&$\Delta[AIC]_{\mathrm{min}}$ & $\Delta[BIC]_{\mathrm{min}}$}
    \startdata
          All observations & & & & & & &\\
         (a) &397&446&-21&...&...&408&311\\
         (a1)&72 &170&-2 &-325 &-276&83&35 \\
         (a2)&102&224&-3 &-295 &-222&113&89\\
         (b) &-11&135&1&... &... &0  &0\\
         (c) &55&201&-1 &...    &...  &66&66\\
         (c new) &31&177&0.1 &...    &...  &42&42\\
         (d) &168&318&-5 &...  &... &179&184\\\hline
         Obs 2 & & & & & & &\\
         (a) &96&100 &-16&   ...&   ...&98&82\\
         (a1)&17&26  &-1 &-79 &-74 &19&8\\
         (a2)&18&30  &-1 &-78 &-71 &21&12\\
         (b) &15&28  &-1 &...   &...   &17&10\\
         (b1)&2 &20  &0.4  &-13  &-8  &5 &2\\
         (b2)&-3 &18  &0.8  &-17 &-10  &0   &0\\
         (c) &19&32  &-1 &...   &...   &21&14\\
         (d) &51&67  &-5 &...   &...   &54&49
    \enddata
    \tablecomments{ (a)\texttt{TBabs*powerlaw}; 
    (a1)\texttt{TBabs*(powerlaw + ezdiskbb)}; (a2)\texttt{TBabs*(powerlaw + tdediscspec)};\\    
    (b)\texttt{TBabs*((1-$f_{\mathrm{part}}$)*powerlaw+$f_{\mathrm{part}}$*XABS*powerlaw)}; \\
    (b1)\texttt{TBabs*((1-$f_{\mathrm{part}}$)*(powerlaw + tdediscspec)+$f_{\mathrm{part}}$*XABS*(powerlaw+tdediscspec)}; \\
    (b2)\texttt{TBabs*((1-$f_{\mathrm{part}}$)*powerlaw+$f_{\mathrm{part}}$*XABS*powerlaw+gsmooth*XEMI)}; \\
    (c)\texttt{TBabs*(relxill + ezdiskbb)}; (d)\texttt{TBabs*XABS*relxill}.\\}
\end{deluxetable*}

\subsection{Model Comparison} \label{sec: AIC}
To decide which model provides the best description of the data, we compute the Akaike information criterion (AIC; \citealt{Akaike_1974}) and Bayesian information criterion (BIC; \citealt{Schwarz_1978}) to assess the goodness of fit. The AIC and BIC for each model are computed by the formulae:
\begin{equation}
    \rm AIC = n \ln{(C / n)} + 2p
\end{equation}
\begin{equation}
    \rm BIC = n \ln{(C / n)} + p\ln{(n)}
\end{equation}
where $C$ is the C-statistic of the fit, $n$ is the number of data bins, and $p$ is the number of free parameters in the model. In general when $C/\nu \geq 1$, the model with the lowest AIC and BIC is the most preferred model. The difference between BIC and AIC is that BIC has a larger penalty term when $n > e^2$, where $e$ is the Euler's number. Columns 1 and 2 of Table \ref{tab:AIC} present the corresponding AIC and BIC values for each model. The upper section of Table \ref{tab:AIC} presents the values across all observations, whereas the lower section is dedicated specifically to Obs 2. We refer $\left | \Delta AIC \right |$ and $ \left |\Delta BIC \right |$ collectively as $\left |\Delta IC \right |$.  As a general rule, $\left |\Delta IC \right | < 2$ suggests both models fit the data at least equally well, $3 < \left | \Delta IC \right | < 7$ provides considerably more support on the model with the lower information criterion (IC), whereas if $\left | \Delta IC \right | > 10$ then the less favored models have either essentially no support or at least those models fail to explain some substantial features of the data \citep{Burnham_2002}. 

Both AIC and BIC consistently indicate that model (b), the partial covering wind absorption model, is the preferred choice with all the observations taken into account. The ordering of the models in descending favorability is as follows: $(b) > (c) \sim (a1) > (a2) > (d) > (a)$. In the comparison of models $(a1)$ and $(c)$, with and without disk reflection, the AIC prefers the relativistic disk reflection over powerlaw, while the BIC prefers the simpler powerlaw plus disk blackbody emission. If only consider Obs 2, the order is $(b2) > (b1) > (b) \sim (a1) > (c) > (a2) > (d) > (a)$. The difference of AIC and BIC between model (b) and model (a1) falls within 2, which indicates that both partial covering wind absorption and disk emission equally fit the Obs 2 data well.

Column 3 of Table \ref{tab:AIC} lists the sigma level of the C-statistics relative to the degrees of freedom, $\sigma_C$, computed by:
\begin{equation}
    \sigma_C = (\nu - C) / \sqrt{2\nu},
\end{equation}
where $\nu$ is the degrees of freedom. Ideally, for a proper model and a perfect fit, the expected value for C-statistics should match the degrees of freedom, with its variance being $2\nu$ when the spectrum has more than $\sim 30$ counts \citep{Kaastra_2017}. A good fit would have a $|\sigma_C|$ value that does not exceed 1. If $\sigma_C$ has a significantly negative value, it suggests that the model fails to capture the main features of the data. Conversely, a substantially positive value of $\sigma_C$ could indicate either an over-fit of the data by the model or excessively large error bars in the data. 

When considering all observations, only model (c) can be considered a good fit according to the value of $\sigma_C$. Model (b), which has the lowest IC values, has a $\sigma_C$ that slightly exceeds 1. This indicates that the partial covering wind model might be too complicated for the data, suggesting a preference for a simpler disk emission plus reflection model. However, focusing exclusively on Obs 2 presents a different scenario. The series of wind models, namely (b), (b1), and (b2), maintain $|\sigma_C|$ values within the acceptable range of 1. In contrast, the disk emission and reflection models (a1), (a2), and (c), have $|\sigma_C|$ values that exceed, but are very close to, 1. Consequently, it is not definitive to state that the wind model is preferred over the disk emission model based solely on this observation. This ambiguity suggests the need for a more nuanced consideration of the models' suitability based on specific observational data. 

Columns 4 - 7 of Table \ref{tab:AIC} lists the AIC and BIC differences that are calculated by:
\begin{equation}
    \Delta[IC]_x = IC_i - IC_x
\end{equation}
where $IC_i$ and $IC_x$ are the information criterion for the current model and for the model ($x$). In Table \ref{tab:AIC}, the subscrip `$p$' means the previous model in the list, `$min$' means the model with minimum IC.

Adding the disk emission in models (a1) and (a2), in contrast to model (a), results in a significant reduction of approximately 300 in both AIC and BIC when all the observations are considered, and about 80 when focusing solely on Obs 2. This substantial reduction in AIC and BIC provides strong evidence supporting the inclusion of the disk emission. Similarly, the addition of either a disk emission or a smoothed wind emission leads to a decrease in both AIC and BIC, by around 10. However, if we compare (b1) and (b2), the difference in AIC and BIC is relatively minor, being 4.1 and 1.9 respectively. This implies some uncertainty in determining the statistically preferred model. As previously discussed in Section \ref{sec:photmodel}, model (b2) may not be physically plausible, which suggests that model (b1), with its combination of disk emission and wind, should be considered the more favorable option among all the models.

On purely statistical grounds, then, when evaluating all observations as a whole, the partial covering wind absorption model (b), emerges as the most suitable for collectively describing the data statistically, but with potential concerns of overfitting. The next alternative is the disk reflection model (c) followed by the basic powerlaw disk emission models (a1) and (a2). When examining Obs 2 individually, the support for model (b) diminishes in comparison to model (a1), with AIC and BIC showing divergent preferences between these models, introducing the ambiguity in selecting superior model fit between model (b) and model (a1).

\section{Discussion and Summary} \label{sec:discussion}
We have analyzed a series of XMM-Newton spectra of AT2021ehb, obtained approximately 300~days after the disruption.  The data achieved a sensitivity that exceeds most observations of TDEs, but do not permit the line-by-line analysis undertaken in TDEs like ASASSN-14li \citep{Miller_2015, Miller_2023} and ASASSN-20qc \citep{Kosec_2023}.  Simple but robust disk plus power-law models (a1 and a2) provide an adequate description of the data, given plausible uncertainties in flux calibrations, backgrounds, and other factors.  The X-ray disk temperatures that result are 50--100\% higher than those found in several other TDEs that permitted basic spectral fits \citep{vanvelzen_2021}; combined with their implied emitting areas, the disk fits imply a black hole mass of $M_{\mathrm{BH}} \simeq 10^{5-6} M_\odot$.  This is clearly at the low end of the black hole mass function, and nominally represents the realization of a long-standing promise of TDEs.   Therefore, we also explored fits with disk reflection components (models c and d) and photoionized winds (models b) that are motivated by phenomena observed in other accreting black hole systems.  These models are not clearly preferred by the data, partly by virtue of the fact that even the simplest implementations are still fairly complex, but they represent viable alternatives and suggest a need for deeper observations of bright TDEs in the future.  The inclusion of plausible disk reflection in the spectral model allows for slightly higher black hole masses.  Separately, allowing for partial covering absorption in a fast ($v = -0.2c$), highly ionized wind may remove the need for an X-ray disk component.  In this section, we discuss the implications of our simple models and complex alternatives.

\subsection{Black Hole Masses and Host Galaxy Properties}
A fundamental prediction from TDE theory is the existence of a maximum black hole mass threshold, beyond which a star can cross the event horizon of a Schwarzschild black hole without experiencing disruption. This threshold is known as the Hills mass \citep{Hills_1975} and is described by the equation:
\begin{equation}
    M_{\mathrm{BH}} < 1.14 \times 10^8 M_{\odot} m_*^{-1/2} r_*^{3/2},
\end{equation}
where $r_* = R_*/R_\odot$ and $m_* = M_*/M_{\odot}$. This suggests that TDEs are more likely to involve lower-mass SMBHs. Higher-mass black holes can only disrupt stars of high mass, a scenario that becomes increasingly rare and challenging. Therefore, TDEs are valuable for probing the lower-mass end of the black hole mass function. 

Currently, the most widely accepted constraints on black hole mass derive from galactic scaling relationships, such as the black hole mass - stellar velocity dispersion relation ($M_\mathrm{BH} - \sigma$; \citealt{Kormendy_Ho_2013}) and the black hole mass - stellar mass relation ($M_\mathrm{BH} - M_{*}$; \citealt{Greene_2020}). \cite{Yao_2022a}  utilized these scaling relationships to estimate the black hole mass of AT2021ehb at approximately $M_{\mathrm{BH}} \sim 10^7 M_{\odot}$ with intrinsic scatter of 0.3 dex for $M - \sigma$ and 0.8 dex for $M_\mathrm{BH} - M_{*}$ relation. However, the scarcity of low-mass black holes in observations raises questions about the accuracy of such mass estimates. More direct methods are needed for estimating the masses of these low-mass black holes.

Our estimation of the black hole mass in AT2021ehb suggests a mass range of  $10^{5-6} M_{\odot}$, which is approximately 1.5 dex lower than the prediction made using galactic scaling relations. This discrepancy could imply either that the galactic relations are not applicable to AT2021ehb, or that our simple models for moderately sensitive data may have failed to capture important details. A black hole mass consistent with scaling relationships would be inferred if the color correction factor (in Equation \ref{eq: Norm}) were 4 times higher than our assumed value of $f = 2.3$. \cite{Merloni_2000} considered the possibility of larger color correction factors up to 3 for accretion disks, including more energy dissipation in the corona. Even for the maximal value of the correction factor, our fits to AT2021ehb would still not suggest a black hole above $10^7~M_{\odot}$. However, our knowledge of disks and coronae in TDEs is still evolving, and future work may determine that higher correction factors are needed, leading to larger masses.

\cite{Yao_2022a} constructed the host galaxy SED of AT2021ehb prior to the disruption, identifying strong Balmer absorption lines but no significant emission in [OII] nor $\mathrm{H}\alpha$ lines, akin to spectra of E+A galaxies \citep{Dressler_Gunn_1983}, with no ongoing star formation that ceased abruptly $\sim$ 1 Gyr ago. TDEs strongly prefer low star formation hosts \citep{Arcavi_2014} and their host galaxies are likely consistent with the same population of galaxies as E+A but older stellar populations \citep{Hammerstein_Gezari_2023}.  Therefore, the host galaxy of AT2021ehb might be a post-starburst galaxy that is evolving from late to early types. This might explain our lower BH mass estimate, as the BHs in late-type AGNs are, on average, about 0.6 dex less massive than early-types \citep{Zhuang_Ho_2023}.

One possible scenario to drive their evolution from late to early-types is that the Super-Eddington accretion can grow under massive BHs and restore them to the local $M_\mathrm{BH} - M_*$ relation of early-type galaxies \citep{Zhuang_Ho_2023}. Similar to the effect of supernova feedback on the evolutionary path of systems with undermassive BHs predicted by \cite{Zhuang_Ho_2023}, the feedback from TDE, possibly due to wind outflows that disrupt the flows that supply material to the accretion disk (will be discussed in \ref{Discussion: Winds}), might lag the growth of the BH behind that of the stars. Later, the BH can catch up with its host after the gas reservoir has become stabilized at higher $M_*$.

The discrepancy in black hole mass estimation could also be attributed to the possibility that models (a1) and (a2) do not adequately represent the data. One major concern is the suppression of X-ray emission due to reprocessing by an ionized medium, which is significant at early epochs with a high ratio of UV/optical blackbody luminosity to X-ray luminosity ($L_{\mathrm{BB}}/L_\mathrm{X}$). The suppression then decreases with time causing the radius to appear to increase to a more reasonable value \citep{Guolo_2023}. In the case of our \xmm\ observations of AT2021ehb, the ratio of $L_{\mathrm{BB}}/L_\mathrm{X}$ is around 1, similar to that of ASASSN-14li \citep{Miller_2015}. Although, the X-ray emission had dropped during the moderately soft state at phase E1, as described in \citet{Yao_2022a}, it remained comparable to the UV/Optical emission, indicating a lesser degree of suppression than the initial soft state where the ratio of $L_{\mathrm{BB}}/L_\mathrm{X}$ is around 200. As the reprocessing medium becomes less ionized over time, the suppression of the X-ray emission diminishes, leading to an apparent increase in the radius and a more accurate black hole mass estimation.

Additionally, the inclusion of disk reflection features in the model (c) led to slightly higher estimates of the black hole mass, which indicates that a more complex model, such as the partial covering wind combined with disk reflection, might yield a more accurate estimation of the black hole's mass. However, distinguishing such a model would require data of higher resolution and sensitivity. As noted by \cite{Parker_2022}, if the spectrum includes hybrid lines resulting from both winds and reflection, it could introduce a significant systematic bias in the estimation of key parameters. Finally, the variance between the black hole mass estimations and the predictions of galactic scaling relationships might hint at a large spin of the black hole.

Alternative approaches have been employed to estimate black hole masses in TDEs. \cite{Mummery_van_2024} developed three techniques for this purpose by using the late-time plateau luminosity observed in optical/UV light curves, using the peak g-band luminosity, and g-band radiated energy. For AT2021ehb, the black hole mass estimations using these methods are $\log M_\mathrm{BH} = 6.76^{+0.52}_{-0.41}$, $5.88 \pm 0.4$, and $6.13 \pm 0.3$, respectively. Notably, our black hole mass estimation largely aligns with the result derived from the peak g-band luminosity.

To further explore the possibility of systematic errors in black hole mass estimation from the disk emitting area, further observations and more detailed modeling, considering the time-dependent changes in the reprocessing medium, are necessary to refine our black hole mass estimates and understand the physical processes at play. Such future work would help determine the relative accuracy of these different methodologies, as it is currently challenging to ascertain the most precise method.  

\subsection{Mass Outflow Rate, Driving Mechanism of the Wind, Mass of the Disrupted Star} \label{Discussion: Winds}
Tests using the AIC and BIC also indicate that photoionization models may be a possible alternative for a disk blackbody. This is notionally consistent with a disk that may have evolved and cooled significantly from its peak flux and temperature, as indicated in Figure \ref{fig: LC} and fits to \swift\ and \nicer\ data by \cite{Yao_2022a}.

The mass of the outflowing gas, considering a covering factor $f_{\mathrm{cov}} = \Omega / 4\pi$ ($\Omega \in [0, 4\pi]$ and $f_{\mathrm{cov}} = 1$ for a spherical shell of gas), can be expressed as:
\begin{equation}
dM_{\mathrm{out}} = 4 \pi f_{\mathrm{cov}} \mu m_p n f_{v} r^2 f_{\mathrm{part}}~dr,
\end{equation}
where $\mu$ is the mean atomic weight (typically $\mu = 1.23$), $m_p$ is the mass of the proton, $n$ is the number density of the absorbing material, $r$ is the distance of the gas from the central source, and $f_{v}$ is the column filling factor along the flow ($f_{v} = 1$ for a homogeneous outflow). We assume a covering factor of $f_{\mathrm{cov}} = 0.5$, based on the argument by \cite{Tombesi_2010} that approximately half of AGNs have ultrafast, high-ionization outflows. The ionization parameter, defined by \cite{Tarter_1969}, is given by:
\begin{equation} \label{eq:xi}
\xi = \frac{L}{n r^2},
\end{equation}
where $L$ is the ionizing luminosity, simplified here as the bolometric luminosity. \cite{Yao_2022a} reported that the X-ray emission brightened at MJD 59547 and dropped twice at around MJD 59590 and MJD 59640. Our \xmm\ observations fall between the first and the second drop of the X-ray luminosity, which is the E1 phase in \cite{Yao_2022a} where X-ray luminosity dropped back to have a similar value as the UV/optical luminosity. By examining Figure 18 of \cite{Yao_2021} and considering that the UV/optical luminosity stays relatively flat, we infer that the bolometric luminosity, the luminosity between 10000 \AA\ to 10 keV, is approximately 6 times the X-ray luminosity within the 0.5--10 keV range. The X-ray luminosity within this energy range is calculated using the formula $L_X = 4 \pi d_L^2 F_X$, where $d_L$ is the luminosity distance of the source and $F_X$ is the X-ray flux for the model obtained using the \texttt{cflux} function in \texttt{xspec} for the best-fit model (a1). The values are listed in Table \ref{table:observationlog}. The averaged bolometric luminosity of AT2021ehb at the time of six epochs of observations suggested by the model is $L \sim 1.5 \times 10^{43}~\mathrm{erg~s^{-1}}$. Thus, the mass outflow rate can be written as: 
\begin{equation}
    \dot M_{\mathrm{out}} = 4 \pi f_{\mathrm{cov}}\mu m_p \frac{L}{\xi}v_{\mathrm{out}}f_v f_{\mathrm{part}}
\end{equation}

Using the best-fit values from model (b), The mass outflow rates per filling factor for six epochs of observations are $\dot M_{\mathrm{out}} / f_v \sim 2.5, 10.9, 3.2, 5.6, 6.4, 2.2 ~\mathrm{M_\odot~yr^{-1}}$. The minimum possible radius of the absorber as the radius at which the observed outflow velocity corresponds to the escape velocity is:
\begin{equation}
    r_\mathrm{min} = 2GM_\mathrm{BH} / v_{\mathrm{out}}^2.
\end{equation}
Using $M_\mathrm{BH} \sim 10^{5.5} M_\odot$, the estimated minimum radius of the absorber is $\log \frac{r_\mathrm{min}}{r_g} = 2.1, 1.6, 1.3, 1.6, 1.6, 2.2$. Manipulating the ionization parameter equation \ref{eq:xi}, and using $N_H = n\Delta r$, we can estimate the minimum possible value of the filling factor $f_{v, \mathrm{min}} = \Delta r / r = N_H \xi r_{\mathrm{min}} / L = 5.5, 0.26, 1.6, 2.0, 1.1, 10.4 \times 10^{-4}$. The minimum possible outflow rate is $\log \dot M_{\mathrm{out, min}} \sim -2.9, -3.6, -3.3, -2.9, -3.1, -2.6~M_\odot/\rm yr$. 

Assuming that the outflow has reached a constant terminal velocity, the kinetic power can be derived as:
\begin{equation}
    \dot E_k = \frac{1}{2} \dot M_{\mathrm{out}} v_{\mathrm{out}}^2
\end{equation}
Hence, the kinetic power is within the range of $10^{42}~\mathrm{erg~s^{-1}} < \dot E_k < 10^{46}~\mathrm{erg~s^{-1}}$. The maximum kinetic energy exceeds the bolometric luminosity by $\sim3$ dex, for a unity volume filling factor. The minimum kinetic energy is as high as $\sim 10\%~L_{bol}$, which exceeds the standards for strong feedback in active galactic nuclei (AGN). Previous studies such as \cite{Di_2005} and \cite{Hopkins_2010} have shown that an outflow with a kinetic power as low as $0.5\%$ of the bolometric luminosity can provide significant feedback on the host galaxy. However, this does not suggest that the observed outflow has helped induce a rapid suppression of star formation and disrupted the cooling flows that supply material to the accretion disk unless the outflows live for a very long time.

For a black hole with a mass of $M_\mathrm{BH} = 10^{5.5} M_\odot$, the Eddington luminosity is calculated as $L_\mathrm{Edd} \equiv 4\pi GM_\mathrm{BH}m_pc / \sigma_T \simeq 3.98 \times 10^{43}~\mathrm{erg~s^{-1}}$, and the Eddington accretion rate is estimated as $\dot M_\mathrm{Edd} \equiv L_\mathrm{Edd}/ \eta c^2 = 0.007~M_\odot / \rm yr$, assuming an accretion radiative efficiency of $\eta \simeq 0.1$ for a standard thin disk \citep{Gruzinov_1998}. Comparing the mass outflow rate with the Eddington rate, we find that $-1.4 < \log{(\dot M_{\mathrm{out}} / \dot M_\mathrm{Edd})} < 3.2$. This indicates that the mass outflow rate is approximately $10\%$ of the Eddington rate for the minimum value of the filling factor, and it can reach as high as $\sim 10^3$ times the Eddington rate for a unity filling factor. The significantly larger mass outflow rate compared to the Eddington accretion rate suggests that most of the matter is being expelled, and only a small fraction is reaching the central black hole at the time of observation. The high mass outflow rates, $\dot M_{\mathrm{out}}  > 0.1 \dot M_\mathrm{Edd}$ suggests that a slim disk \citep{Sadowski_2009} model is a good description of TDE disks \citep{Wen_2020} when they are accreting at a high fraction of Eddington limit. In replacement of the \texttt{ezdiskbb} in model (a1), we fit the data with \texttt{diskpbb}, which is a multi-temperature blackbody disk model with an exponential radial dependence of the local disk temperature. The standard disk model is recovered when the exponent of the radial temperature dependence, denoted as $p$, is set to 0.75. The data analysis reveals that the 1-sigma values of the exponent parameter $p$ span a large range from 0.6 to 1, suggesting that the observational data alone cannot definitively determine whether the accretion disk is slim or not.

At the time of the \xmm\ observation, approximately 300 days after the peak, AT2021ehb exhibits a shallower $F \propto t^{-5/12}$ power-law decay according to the UV light curve of \swift\ in Figure \ref{fig: LC}. This suggests that the fallback rate has dropped to sub-Eddington values. Based on the formula in equation 6 from \cite{Lodato_2011}, the peak fallback rate for disrupting a solar-like star is estimated to be $\dot M_p = 5~M_\odot /\mathrm{yr}$ at the time of disruption.  Following roughly $t_{\mathrm{min}} = 23$ days post-disruption, the material starts coming back to pericentre at a rate $\dot M_{\mathrm{fb}}$. Assuming a classic dependence of $t^{-5/3}$ for the fallback rate, the estimated fallback rate at the time of the observation, roughly 300 days after the disruption, is roughly $\dot M_{\mathrm{fb}} = 0.07~M_\odot /\rm yr$. This fallback rate is as large as 10 times the Eddington rate assuming a disruption of a solar mass star, suggesting that the disrupted star has a mass lower than the Sun's, or the black hole's mass is greater than our current estimate, or that there is a massive outflow. If we assume $\dot M_{\mathrm{fb}} \sim \dot M_\mathrm{out}$, the volume filling factor should be $f_v = 2.8 \times 10^{-4}$, which is close to the lower end of the previous estimation where we assumed the outflow occurs at the minimum possible radius from the black hole.

In case the mass of the black hole is higher, we evaluate the changes in the result we have discussed here. For a higher black hole mass $\sim 10^7 M_{\odot}$, the Eddington limit is higher. The mass outflow rate is unchanged for a unity volume filling factor. The minimum radius increased by $\sim 100$ resulting in a decrease of the minimum possible value of the filling factor, hence, compensating the increase of the Eddington mass outflow and resulting in a nearly unchanged lower limit of $\log (\dot M_{\mathrm{out}}/\dot M_\mathrm{Edd})$. The upper limit is decreased by $\sim 1.5$ dex. But the mass outflow rate is still as high as $\sim 10^{1.5}$ times the Eddington rate. The fallback rate gets increased to about $0.2 M_{\odot} / \mathrm{yr}$, the same as the Eddington rate assuming a disruption of a solar mass star, suggesting a higher mass of the disrupted star but still should be lower than the Sun's. The volume filling factor estimated by $\dot M_{\mathrm{fb}} \sim \dot M_\mathrm{out}$ still matches that evaluated by the minimum radius.

We can have more insight into the acceleration mechanisms of the winds by comparing the outflow and radiation momentum rates. The outflow momentum rate can be expressed as:
\begin{equation}
    \dot P_{\mathrm{out}} = \dot M_{\mathrm{out}} v_{\mathrm{out}},
\end{equation}
while the momentum flux of the radiation fields is defined as:
\begin{equation}
    \dot P_{\mathrm{rad}} = \frac{L_{bol}}{c}.
\end{equation}
If we consider $\dot M_{\mathrm{out}}$ as the minimum estimated values for $M_\mathrm{BH} = 10^{5.5} M_\odot$, the ratio of $\frac{P_{\mathrm{out}}}{P_{\mathrm{rad}}} = 0.64, 0.24, 0.60, 0.60, 0.96, 0.93$ for all six epochs of observations. The values agree with previous studies, which have indicated that the average ratio for typical UFOs is consistent with unity, while Warm Absorbers (WAs) exhibit significantly lower ratios \citep{Tombesi_2013, Laha_2016}. However, for the higher mass of the black hole $M_\mathrm{BH} = 10^{7} M_\odot$, the ratio increased by $\sim 100$, either challenging the UFO scenario or suggesting that the mass of the black hole should be lower. The electron optical depth to Compton scattering $\tau_e$ can be approximated as $\tau_e \sim \sigma_T N_H \sim 0.3$ for a column density of the wind at $N_H \sim 5\times10^{23}~\mathrm{cm^{-2}}$. This value is lower than $\sim$ unity expected from $\dot P_{\mathrm{out}} \sim \dot P_\mathrm{rad}$ \citep{Tombesi_2013}. This indicates that the outflow in our model is less influenced by Compton scattering, and the continuum radiation is not the primary driver of acceleration for the flows at the moment of observation \citep{Gofford_2015, Laha_2016}.

Furthermore, the wind is unlikely to be driven by radiation pressure due to lines, since a highly ionized cloud ($\log{\xi} > 3$) would lack enough UV transitions to invoke radiation pressure \citep{Dannen_2019}.  It is possible, though, that other parts of the wind are less ionized, making radiation pressure effective, and that those parts of the wind drag out the more ionized gas that we have detected. Magnetic driving may be a plausible means of driving the fast wind in AT2021ehb, based on studies of UFOs in AGN \citep{Fukumura_2010}.  If a standard Shakura-Sunyaev disk is present, the corresponding magnetic flux can help drive the wind. According to \cite{Fukumura_2022}, the outflows driven by magnetohydrodynamics have blueshifted tails while those driven by radiation have an asymmetric line shape of an extended red wing. Unfortunately, our current data set lacks the resolution necessary to identify these distinct features. The capability to distinguish between these characteristics, alongside the opportunity to study the evolution of the flow, is anticipated in high-resolution calorimeter data from missions such as XRISM/Resolve \citep{Tashiro_2020} and upcoming Athena/XIFU \citep{Barret_2023}. 

Our findings highlight the importance of considering photoionization absorption winds and their impact on the observed spectra. The recent launch of XRISM \citep{Tashiro_2020} may further enhance our understanding of TDEs by providing high-resolution spectroscopy and a comprehensive view of the X-ray emission. These observations will allow for detailed measurements of line profiles, line ratios, kinematics, and abundances, enabling us to probe the properties of the emitting gas and gain deeper insights into the dynamics and physical conditions of accretion and outflow processes. We suggest continued observations of AT2021ehb in the later stages of its decay to monitor the accretion flow's evolution towards a stable and enduring configuration. With ongoing advancements in observational capabilities, we anticipate uncovering more insights into the intriguing phenomena of TDEs and their role in shaping the evolution of galaxies.

\begin{acknowledgments}
We thank the XMM-Newton Principal Investigator, Norbert Schartel, and the planning and scheduling scientists for making these observations possible.  We are also grateful to Brad Cenko and the Swift planning and scheduling teams for monitoring this TDE.  We acknowledge helpful conversations with Sean Johnson, Andrew Mummery, and Richard Mushotzky.
\end{acknowledgments}


\bibliography{main}{}
\bibliographystyle{aasjournal}

\end{document}